# Algebraic Properties for Selector Functions[1]

Lane A. Hemaspaandra[2]
Department of Computer Science
University of Rochester
Rochester, NY 14627-0226, USA

Harald Hempel[3]
Institut für Informatik
Friedrich-Schiller-Universität Jena
D-07743 Jena, Germany

Arfst Nickelsen[4]
Fachbereich Informatik
Technische Universität Berlin
D-10587 Berlin, Germany

January 11, 2005[5]

[1] Earlier versions of some parts of this paper were presented at the COCOON 2001 conference.
[2] `lane@cs.rochester.edu`. Supported in part by grants NSF-CCR-9322513 and NSF-INT-9815095/ DAAD-315-PPP-gü-ab.
[3] `hempel@informatik.uni-jena.de`. Supported in part by grant NSF-INT-9815095/DAAD-315-PPP-gü-ab. Work done while visiting the University of Rochester under a NATO Postdoctoral Science Fellowship from the Deutscher Akademischer Austauschdienst's "Gemeinsames Hochschulsonderprogramm III von Bund und Ländern" program.
[4] `nicke@cs.tu-berlin.de`.
[5] This is identical to the January 7, 2004 revision of Univ. of Rochester Comp. Sci. Dept. Technical Report TR-2002-778.

**Abstract**

The *nondeterministic* advice complexity of the P-selective sets is known to be exactly linear. Regarding the *deterministic* advice complexity of the P-selective sets—i.e., the amount of Karp–Lipton advice needed for polynomial-time machines to recognize them in general—the best current upper bound is quadratic [Ko83] and the best current lower bound is linear [HT96].

We prove that every associatively P-selective set is commutatively, associatively P-selective. Using this, we establish an algebraic sufficient condition for the P-selective sets to have a linear upper bound (which thus would match the existing lower bound) on their deterministic advice complexity: If all P-selective sets are associatively P-selective then the deterministic advice complexity of the P-selective sets is linear. The weakest previously known sufficient condition was P = NP.

We also establish related results for algebraic properties of, and advice complexity of, the nondeterministically selective sets.

# 1 Introduction

Selman [Sel79, Sel81, Sel82a, Sel82b] defined the P-selective sets about twenty years ago. In addition to being of interest in their own right, they have recently had some surprising applications. For example, selectivity is a powerful tool in the study of search versus decision problems [HNOS96a], and nondeterministic generalizations of selectivity are the key tools used to show that even NP machines cannot uniquely refine satisfying assignments unless the polynomial hierarchy collapses [HNOS96b], that even weaker refinements are also precluded unless the polynomial hierarchy collapses [Ogi96, NRRS98], and that many cardinality types of nondeterministic function classes cannot collapse unless the polynomial hierarchy collapses [HOW02].

**Definition 1.1 ([Sel79])** *A set $B$ is* P-*selective if and only if there is a (total) polynomial-time function $f : \Sigma^* \times \Sigma^* \to \Sigma^*$ such that*

1. *$(\forall x, y)[f(x, y) = x \ \lor \ f(x, y) = y]$, and*
2. *$(\forall x, y)[(x \in B \ \lor \ y \in B) \implies f(x, y) \in B]$.*

*We call such a function $f$ a* P-*selector function for $B$.*

That is, a set $B$ is P-selective if there is a polynomial-time function $f$ that given any two strings always chooses one of them, and $f$ does this in such a way that if exactly one of the strings is in $B$ then the P-selector function chooses that string. $f(x, y)$ is often described, very informally, as choosing one of $x$ or $y$ that is more likely to be in the set, though a more accurate description would be that $f$ chooses one of $x$ or $y$ that is logically no less likely then the other to be in the set.

The P-selective sets have been extensively studied, and much about them is well understood (see the recent book [HT03] about selectivity theory). Though some P-selective sets are very complex—highly undecidable—the P-selective sets nonetheless have a broad range of structural simplicity properties. Most crucially in terms of the study in this paper, Ko [Ko83] showed that they have low nonuniform complexity (P-sel $\subseteq$ P$/\mathcal{O}(n^2)$). A few among the many other simplicity results known to hold are: Ko and Schöning [KS85] showed that all P-selective sets in NP are in the second level of the low hierarchy of Schöning [Sch83], and Allender and Hemachandra [AH92] showed that the Ko–Schöning result is the strongest lowness result for P-sel that holds with respect to all oracles; a long line of work starting with Selman [Sel79] and Toda [Tod91] (see also [Siv99] and the citations therein) has shown that no P-selective set can be NP-hard under $\leq^p_m$ or various other reductions unless P = NP; Naik and Selman [NS99] have shown that no P-selective set can be truth-table-hard for NP unless certain (intuitively unlikely) containments hold in the relationship between adaptive and nonadaptive queries to NP; and as a consequence of the work of Ko [Ko83] and Cai [Cai01] no P-selective set can be truth-table-hard for NP (or even Turing-hard for NP) unless the polynomial hierarchy collapses to S$_2$, where S$_2$ is the symmetric alternation class of Canetti [Can96] and Russell and Sundaram [RS98]. Note that S$_2 \subseteq$ ZPP$^{\text{NP}} \subseteq$ NP$^{\text{NP}}$ [Cai01], where ZPP as usual denotes expected polynomial time, so this is a very dramatic collapse.

In this paper, we show that sets having P- or NP-selector functions with nice algebraic properties have simplicity properties far beyond those known for general P-selective or NP-selective sets. More generally, we study the class of languages one obtains from P- and NP-selector functions having certain algebraic properties, and the possibility of obtaining such algebraically nice P- or NP-selector functions.

In particular, Section 4 shows that any P-selective (NP-selective) set having an associative P-selector (NP-selector) function also has a commutative, associative P-selector (NP-selector) function. Section 5 shows that sets having an associative P-selector can be accepted by deterministic advice interpreters using only a linear amount of advice. In contrast, the best upper bound on the deterministic advice complexity of the P-selective sets is the quadratic bound obtained by Ko [Ko83].

Our result provides a new sufficient condition—all P-selective sets are associatively P-selective—for all P-selective sets having linear deterministic advice; the weakest previously known sufficient condition was the quite demanding assumption that P = NP.

Section 6 shows that associatively NP-selective sets with weak census functions cannot be coNP-immune. Section 7 establishes a structural sufficient condition for all P-selective sets being associatively, commutatively P-selective, and proves that if all NPMV-selective sets have associative NPMV-selector functions then the polynomial hierarchy collapses.

## 2 Definitions

Commutativity and associativity are concepts often associated with single-valued total functions. However, as we will discuss soon, these concepts can be naturally applied to multivalued and/or partial functions (see, e.g., [RS97,HR99]). For a 2-ary multivalued function $f$, let $set\text{-}f(x,y)$ denote the set of values of $f$ on input $(x,y)$.[1] $f(x,y)$ is undefined if and only if $set\text{-}f(x,y) = \emptyset$. We extend this notation to single-valued functions in the obvious way. For a 2-ary single-valued function $f$ let $set\text{-}f(x,y) = \{f(x,y)\}$ if $f(x,y)$ is defined and let $set\text{-}f(x,y) = \emptyset$ otherwise. For a (single- or multivalued) function $f$, a set $A$, and a string $y$, define $set\text{-}f(A,y) = \bigcup_{a \in A} set\text{-}f(a,y)$ and $set\text{-}f(y,A) = \bigcup_{a \in A} set\text{-}f(y,a)$. A 2-ary function $f$ is called *total on a set $B$* if and only if $f(x,y)$ is defined for all $x, y \in B$. A function is called total if it is total on $\Sigma^*$.

A total 2-ary (single- or multivalued) function $f$ is *associative on a set $B$* if and only if $set\text{-}f(a, set\text{-}f(b,c)) = set\text{-}f(set\text{-}f(a,b), c)$ holds for all $a, b, c \in B$.[2] (Note that for total single-valued functions this is equivalent to saying that $f(a, f(b,c)) = f(f(a,b), c)$ holds for all $a, b, c \in B$.) We say a total function $f$ is *associative* if $f$ is associative on $\Sigma^*$. A total 2-ary (single- or multivalued) function $f$ is *associative at each length* if $f$ is associative on $\Sigma^n$ for each $n$ (where $\Sigma^n$ is the set of words of length $n$). A 2-ary function $f$ (partial or total, single- or multivalued) is *commutative* if and only if $set\text{-}f(a,b) = set\text{-}f(b,a)$ holds for all $a$ and $b$. (For total single-valued functions this is equivalent to saying that $f(a,b) = f(b,a)$ holds for all $a, b$.)

For partial functions the literature distinguishes two notions of associativity, namely, (strong) associativity and weak associativity [HR99,RS97]. A partial 2-ary function $f$ is called associative (or strongly associative) if and only if $set\text{-}f(a, set\text{-}f(b,c)) = set\text{-}f(set\text{-}f(a,b), c)$ holds for all $a$, $b$, and $c$. A partial 2-ary function $f$ is called weakly associative if and only if for all $a$, $b$, and $c$, either (a) $set\text{-}f(a, set\text{-}f(b,c)) = \emptyset$ or $set\text{-}f(set\text{-}f(a,b), c) = \emptyset$ or (b) $set\text{-}f(a, set\text{-}f(b,c)) = set\text{-}f(set\text{-}f(a,b), c)$. In other words, if any of the four applications of $f$ are undefined, weak associativity "forgives" the equality requirement. The two approaches to associativity for functions correspond to the two notions of equality for partial functions, which date back to work of Kleene [Kle52].

---

[1]Formally, one should speak of relations rather than functions. However the longstanding convention in computer science is to refer to such objects as "multivalued functions" (see [BLS84,BLS85,Sel96]). Similarly, the notation "$set\text{-}f(...)$" is also standard in the literature on multivalued functions, and we follow it. Though this notation makes equations a bit longer and less beautiful (than in the reasonable alternative, which some people prefer, of viewing $f$ as standing for $set\text{-}f$), it does avoid some confusion. For example, suppose $f$ is undefined when its first argument is $x_0$ and its second argument is $y_0$. Then we would say that $f(x_0, y_0)$ is undefined. In contrast, $set\text{-}f(x_0, y_0)$ certainly *is* defined, and in particular $set\text{-}f(x_0, y_0) = \emptyset$. Readers particularly interested in the history, types, and notations involving multivalued functions may wish to look at the interesting papers of Book, Long, and Selman mentioned earlier in this footnote.

[2]As it stands, the definition allows the possibility that some of the values or all of the values in $set\text{-}f(b,c)$, $set\text{-}f(a,b)$, $set\text{-}f(a, set\text{-}f(b,c))$, and $set\text{-}f(set\text{-}f(a,b), c)$ might not belong to $B$. However, since we will typically be applying this to total, *self-contained* functions, the four just-mentioned sets will each be nonempty subsets of $B$.

Note that both of these definitions, for total functions, exactly coincide with the associativity definition for total functions given earlier.

**Definition 2.1 ([HHN$^+$95, HNOS96b])** *For any class of (partial or total, single- or multivalued) functions $\mathcal{F}$, we say a set $B$ is $\mathcal{F}$-selective exactly if there is a function $f \in \mathcal{F}$ such that*

1. $(\forall x, y)[\text{set-}f(x, y) \subseteq \{x, y\}]$, *and*

2. $(\forall x, y)[\{x, y\} \cap B \neq \emptyset \implies (\text{set-}f(x, y) \neq \emptyset \wedge \text{set-}f(x, y) \subseteq B)]$.

*A function $f$ with property 1 is called a self-contained function. A function with properties 1 and 2 is called an $\mathcal{F}$-selector function for $B$ (when $B$ is clear from context or unimportant we will simply speak of an $\mathcal{F}$-selector function, and when $\mathcal{F}$ is clear from context or unimportant we will simply speak of a selector function (for $B$)).*

This definition requires a selector function for a set $B$ to be defined whenever at least one input is in $B$. In other words, partial selector functions can only be undefined if both inputs are in $\overline{B}$. Observe that for any *total* self-contained function $f$ we have that for any strings $a$, $b$, and $c$, $\text{set-}f(a,b) \cap \text{set-}f(a,c) \subseteq \text{set-}f(a, \text{set-}f(b,c)) \subseteq \text{set-}f(a,b) \cup \text{set-}f(a,c)$ and $\text{set-}f(a,c) \cap \text{set-}f(b,c) \subseteq \text{set-}f(\text{set-}f(a,b),c) \subseteq \text{set-}f(a,c) \cup \text{set-}f(b,c)$.

We will use the following notational shorthands for the classes that we will study. (The ordering of "A" and "C" changes to avoid confusion with the existing class "AC," and to make clear that the "$\ell$" modifies just the "A.")

**Definition 2.2** *For any class of (partial or total, single- or multivalued) functions $\mathcal{F}$:*

1. $\mathcal{F}\text{-sel} = \{B \mid B \text{ is } \mathcal{F}\text{-selective}\}$.

2. $\text{A-}\mathcal{F}\text{-sel} = \{B \mid B \text{ is } \mathcal{F}\text{-selective via an associative } \mathcal{F}\text{-selector function}\}$.

3. $\text{C-}\mathcal{F}\text{-sel} = \{B \mid B \text{ is } \mathcal{F}\text{-selective via a commutative } \mathcal{F}\text{-selector function}\}$.

4. $\text{CA-}\mathcal{F}\text{-sel} = \{B \mid B \text{ is } \mathcal{F}\text{-selective via a commutative, associative } \mathcal{F}\text{-selector function}\}$.

5. $\text{A}_\ell\text{-}\mathcal{F}\text{-sel} = \{B \mid B \text{ is } \mathcal{F}\text{-selective via an } \mathcal{F}\text{-selector function that is associative at each length}\}$.

6. $\text{A}_\ell\text{C-}\mathcal{F}\text{-sel} = \{B \mid B \text{ is } \mathcal{F}\text{-selective via a commutative } \mathcal{F}\text{-selector function that is associative at each length}\}$.

FP will denote the (possibly partial) polynomial-time computable functions. $\text{FP}_t$ will denote those functions in FP that are total. (In the literature, some authors use the notation "FP" to denote what we in this paper denote $\text{FP}_t$. To avoid confusion, we urge the reader to note carefully that throughout this paper FP (and also the function classes NPSV and NPMV to soon be defined) do not require totality except when subscripted with a "$t$", i.e., $\text{FP}_t$, $\text{NPSV}_t$, and $\text{NPMV}_t$.)

For any set $C$, $\text{FP}^C$ will be the class of all (possibly partial) functions that are computable in polynomial time with the help of oracle $C$. For any class $\mathcal{C}$, $\text{FP}^\mathcal{C} = \{g \mid (\exists C \in \mathcal{C})[g \in \text{FP}^C]\}$. $\text{FP}_t{}^C$ and $\text{FP}_t{}^\mathcal{C}$ are defined analogously.

FP and $\text{FP}_t$ functions each have some fixed arity for their domain, and their codomain is of arity one and is $\Sigma^*$. That is, FP and $\text{FP}_t$ are, respectively, potentially-partial and per-force-total functions, each of type $\Sigma^* \to \Sigma^*$, or of type $\Sigma^* \times \Sigma^* \to \Sigma^*$, or of type $\Sigma^* \times \Sigma^* \times \Sigma^* \to \Sigma^*$, etc. However, for selector functions, which are the main focus of this paper, the (input) arity is always two—selector functions by their nature are two-argument functions. So, throughout this paper, whenever we speak of selector-type functions (or related functions that by their nature or the way we use them are clearly 2-ary), we will simply write expressions such as "there is an $f \in \text{FP}$" when in fact we mean "there is a 2-ary function $f \in \text{FP}$."

We will follow the same convention regarding the nondeterministic function classes NPSV, $\text{NPSV}_t$, NPMV, and $\text{NPMV}_t$. These classes in general are meaningful for 1-ary functions, for 2-ary functions, etc. However, since this paper focuses on selector functions, which are always 2-ary, we will in effect silently treat the classes as if they are over 2-ary functions. So, for example, in the following paragraph, our machines have two arguments. (The case of $k$-ary, $k \neq 2$, functions belonging to these classes is identically defined except with $k$ arguments.) And throughout this paper we simply say, for our nondeterministic function classes, things such as "there is an $f \in \text{NPSV}$" when in fact we mean "there is a 2-ary function $f \in \text{NPSV}$."

In the literature (and in the present paper), $\text{FP}_t$-sel is often denoted P-sel (note that Definition 2.2 for $\mathcal{F} = \text{FP}_t$ yields the same class of P-selective sets as defined in Definition 1.1), and $\text{NPSV}_t$-sel is often denoted NP-sel. (Both $\text{FP}_t$-sel and $\text{NPSV}_t$-sel in fact, by the convention mentioned in the previous paragraph, actually are intended to refer to functions of type $\Sigma^* \times \Sigma^* \to \Sigma^*$ from $\text{FP}_t$ and $\text{NPSV}_t$, respectively.)

We now define the standard nondeterministic function classes [BLS84, BLS85]. For any nondeterministic polynomial-time Turing machine $M$ (NPTM for short) taking two strings as its input and for any two strings $x$ and $y$, let $out_M(x, y)$ denote the set of all strings that are output on some accepting path of $M(x, y)$, by which we mean $M$ with $x$ as its first argument and $y$ as its second argument. Note that we view the collection of outputs as a set, not as a multiset (if one path outputs 101 and seven paths output 1100, the set of outputs is simply $\{101, 1100\}$), and note also that the functions computed by such machines may in some cases be partial functions, and may in some cases be multivalued functions. NPSV is the class of all single-valued functions $f$ such that there exists an NPTM $M$ such that, for all $x, y \in \Sigma^*$, $out_M(x, y) = \{f(x, y)\}$ if $f(x, y)$ is defined and $out_M(x, y) = \emptyset$ if $f(x, y)$ is undefined. NPMV is the class of all functions $f$ such that there exists an NPTM $M$ such that, for all $x, y \in \Sigma^*$, $out_M(x, y) = set\text{-}f(x, y)$. $\text{NPSV}_t$ and $\text{NPMV}_t$ denote the set of all total NPSV and NPMV functions, respectively.

As to levels of the polynomial hierarchy [MS72, Sto76] beyond NP, as is standard we will use $\Sigma_2^p$ to denote the class of all languages that can be accepted by nondeterministic polynomial-time Turing machines with the help of an NP oracle, i.e., $\Sigma_2^p = \text{NP}^{\text{NP}}$. And as is standard we will use $\Sigma_3^p$ to denote the class of all languages that can be accepted by nondeterministic polynomial-time Turing machines with the help of a $\Sigma_2^p$ oracle, i.e., $\Sigma_3^p = \text{NP}^{\Sigma_2^p} = \text{NP}^{\text{NP}^{\text{NP}}}$.

Just as P-sel and NP-sel are standard notational conventions when $\text{FP}_t$-sel and $\text{NPSV}_t$-sel are actually meant (see [Sel79] and [HHN$^+$95]), we similarly write A-P-sel and A-NP-sel as shorthands for A-$\text{FP}_t$-sel and A-$\text{NPSV}_t$-sel, respectively, and we do likewise for the other four prefixes C-, CA-, $\text{A}_\ell$-, and $\text{A}_\ell\text{C}$-.

From the definitions,
$$\text{CA-P-sel} \begin{array}{c} \subseteq \text{A-P-sel} \subseteq \\ \subseteq \text{A}_\ell\text{C-P-sel} \subseteq \end{array} \text{A}_\ell\text{-P-sel} \subseteq \text{P-sel}$$
and
$$\text{A}_\ell\text{C-P-sel} \subseteq \text{C-P-sel} \subseteq \text{P-sel}.$$

Exactly the same inclusion relations hold for the various subtypes of NP-selective and of $\text{NPMV}_t$-selective sets.

The following facts are well-known.

**Fact 2.3** *1.* FP-sel = P-sel = C-P-sel.

2. NP-sel = C-NP-sel.

   3. NPSV-sel = C-NPSV-sel.

   4. $\text{NPMV}_t$-sel = $\text{C-NPMV}_t$-sel.

   5. NPMV-sel = C-NPMV-sel.

Regarding the first equality of part 1 of Fact 2.3, recall that any partial selector function can only be undefined if both of its inputs are not in the selected set. Of course one can alter a deterministic polynomial-time Turing machine computing a partial selector function $f$ for a set $A$ in such a way that it outputs any one of the two inputs, say the lexicographically larger one, in the case that it "plans" to output nothing. The total function computed by the altered Turing machine is also a selector function for $A$. The other facts hold since if a set $B$ is $\mathcal{F}$-selective via an $\mathcal{F}$-selector function $f$ (with $\mathcal{F} \in \{\text{FP}_t, \text{NPSV}_t, \text{NPSV}, \text{NPMV}_t, \text{NPMV}\}$) then the function $f'$ defined by

$$f'(x,y) = f(\min(x,y), \max(x,y)) \qquad (*)$$

is a commutative $\mathcal{F}$-selector function for $B$.

Advice classes capture the information content of sets with respect to some complexity class of decoder sets.

**Definition 2.4 ([KL80])**   1. For a complexity class $\mathcal{C}$ and a function $f : \mathbb{N} \to \mathbb{N}$, a set $B$ is in $\mathcal{C}/f(n)$ if and only if there exist a set $C \in \mathcal{C}$ and a function $h : 1^* \to \Sigma^*$ such that, for all $x \in \Sigma^*$,

   (a) $|h(1^{|x|})| = f(|x|)$, and
   (b) $x \in B \iff \langle x, h(1^{|x|}) \rangle \in C$.

   2. For any complexity class $\mathcal{C}$ and any function class $\mathcal{F}$, $\mathcal{C}/\mathcal{F} = \bigcup_{f \in \mathcal{F}} \mathcal{C}/f(n)$.

In this paper, we are particularly interested in $\text{P}/\mathcal{O}(n)$, $\text{P}/n+1$, $\text{P}/n$, $(\text{NP} \cap \text{coNP})/n+1$, and $\text{NP}/n+1$. Also, poly denotes the class of polynomials mapping from $\mathbb{N}$ to $\mathbb{N}$, and we will use later classes of the form $\mathcal{C}/\text{poly}$.

For each set of strings $A$ and each natural number $n$, let $A^{=n}$ and $A^{\leq n}$ denote all strings in $A$ of length exactly $n$ and of length up to and including $n$, respectively. Let $\mathcal{F}$ be any function class. A set $B$ is $\mathcal{F}$-printable if there is a function $g \in \mathcal{F}$ such that, for all $n \in \mathbb{N}$, $g(1^n)$ outputs $B^{\leq n}$ in some fixed standard way of encoding sets as strings [HY84]. That is, there is a function from $\mathcal{F}$ that finds all elements in the set up to a given length. By tradition, P-printable denotes the $\text{FP}_t$-printable sets.

Consider a relation $R$ over the universe $U$, i.e., $R \subseteq U \times U$. We say that $R$ is reflexive if and only if $(\forall a \in U)[(a,a) \in R]$. $R$ is said to be antisymmetric if and only if $(\forall a, b \in U)[((a,b) \in R \land (b,a) \in R) \implies a = b]$. We say that $R$ is a *transitive relation* if and only if $(\forall a, b, c \in U)[((a,b) \in R \land (b,c) \in R) \implies (a,c) \in R]$. For each $a, b \in U$, $a$ and $b$ are called comparable with respect to $R$ if $(a,b) \in R \lor (b,a) \in R$. $R$ is called a partial order on $U$ if and only if $R$ is reflexive, antisymmetric, and transitive. (Informally, order mirrors the relation "less than or equal to.") $R$ is called a linear order (or total order, or, for short, we may sometimes simply write order) on $U$ if and only if $R$ is a partial order on $U$ and each pair of elements from $U$ is comparable with respect to $R$. Given a set $U$ and a relation $\leq$ that is a partial order on $U$, (a) an element $x \in U$ is said to be maximal if and only if $(\forall y \in U)[x \leq y \implies x = y]$; (b) an element $x \in U$ is said to be minimal if and only if $(\forall y \in U)[y \leq x \implies x = y]$. If a maximal (minimal) element exists in a linear order it is unique and can be referred to as maximum (minimum). (For a review of the basics of orders see, e.g., [Cam94, Chapter 12].)

As is standard in graph theory, in a directed graph (digraph for short) $G = (V, E)$ we have that $V$ is a finite set and $E \subseteq V \times V$. A directed graph $G = (V, E)$ is called a *tournament* if and only if (a) for all $u, v \in V$ such that $u \neq v$, exactly one of the edges $(u, v)$ and $(v, u)$ belongs to $E$, and (b) for all $u \in V$, it holds that $(u, u) \notin E$. We will in this paper slightly modify this notion by adding self-loops at each node, that is, a directed graph $G = (V, E)$ is called a *self-looped tournament* (s-tournament, for short) if and only if (a) for all $u, v \in V$ such that $u \neq v$, exactly one of the edges $(u, v)$ and $(v, u)$ belongs to $E$, and (b) for all $u \in V$, it holds that $(u, u) \in E$. (If we were dealing only with tournament-type graphs this modification would not be useful. However, in this paper we study even multivalued selector functions, and the natural graphs that these induce are of a different flavor, and in particular when they are transitive and contain cycles will have self-loops along all cycles even if we do not build self-loops into the way we induce these graphs. Thus, to allow the similarity of the relationship—which we will later see—between associativity of selector functions and transitivity of selector-induced digraphs to be as clear as possible, we use s-tournaments rather than tournaments.)

A directed graph $G = (V, E)$ is called a *complete digraph* (or synonymously a *weak clique*) if and only if, for all $u, v \in V$, *at least* one of the edges $(u, v)$ and $(v, u)$ belongs to $E$. A directed graph $G = (V, E)$ is called a *strong clique* if and only if, for all $u, v \in V$, the edge $(u, v)$ belongs to $E$. (Note that both complete digraphs and strong cliques by these definitions must have every self-loop.) A subgraph $G'$ of $G$ is called a maximal strong clique if $G'$ is a strong clique and there is no subgraph $G''$ in $G$ that is a strong clique and contains $G'$ as a proper subgraph.

As is standard, given a directed graph $G = (V, E)$, a directed graph $G' = (V', E')$ is said to be an *induced subgraph of $G$* exactly if $V' \subseteq V$ and $E' = \{(a, b) \mid a \in V' \land b \in V' \land (a, b) \in E\}$.

## 3   Commutative Selectors and Digraphs

Parts 1, 2, and 4 of Fact 2.3 establish that without loss of generality one can assume P- (NP- or NPMV$_t$-) selectors to be commutative. Or, to be more precise, they make clear that restricting oneself to commutative selector functions in these contexts in no way shrinks the class of sets obtained. This observation enables one to use results from graph theory, in particular, from tournament theory, to pinpoint the advice complexity of selective sets.

Below we would like to make the link between commutative selector functions and directed graphs (in special cases, s-tournaments) more explicit and prove some results regarding this link.

Since our focus will be on self-contained associative functions, the following easy observation will be quite useful in the upcoming proofs. Note that the functions $g$ spoken of in Proposition 3.1 may potentially be nontotal and/or multivalued.

**Proposition 3.1** *Let $B \subseteq \Sigma^*$ and let $g$ be any commutative self-contained function that is total on $B$. Then the following are equivalent:*

1. *$g$ is associative on $B$.*

2. *$(\forall a, b, c \in B)[\text{set-}g(a, \text{set-}g(b, c)) = \text{set-}g(b, \text{set-}g(a, c)) = \text{set-}g(c, \text{set-}g(a, b))]$.*

3. *$(\forall a, b, c \in B : \|\{a, b, c\}\| = 3)[\text{set-}g(a, \text{set-}g(b, c)) = \text{set-}g(b, \text{set-}g(a, c)) = \text{set-}g(c, \text{set-}g(a, b))]$.*

**Proof:** Let $g$ be a commutative self-contained function that is total on $B$, $B \subseteq \Sigma^*$. Note that part 1 in light of $g$'s commutativity implies part 2 and that part 2 implies part 3. We will complete the proof by showing that part 3 implies part 2 and that part 2 implies part 1.

Suppose that part 2 holds, i.e., $(\forall a,b,c \in B)[\textit{set-g}(a,\textit{set-g}(b,c)) = \textit{set-g}(b,\textit{set-g}(a,c)) = \textit{set-g}(c,\textit{set-g}(a,b))]$. Let $x,y,z \in B$. By our assumption we have $\textit{set-g}(x,\textit{set-g}(y,z)) = \textit{set-g}(z,\textit{set-g}(x,y))$. Since $g$ is total on $B$ and commutative we have $\textit{set-g}(z,\textit{set-g}(x,y)) = \textit{set-g}(\textit{set-g}(x,y),z)$ and thus $\textit{set-g}(x,\textit{set-g}(y,z)) = \textit{set-g}(\textit{set-g}(x,y),z)$. This by definition shows that $g$ is associative.

Now suppose that part 3 holds, i.e., $(\forall a,b,c \in B : ||\{a,b,c\}|| = 3)[\textit{set-g}(a,\textit{set-g}(b,c)) = \textit{set-g}(b,\textit{set-g}(a,c)) = \textit{set-g}(c,\textit{set-g}(a,b))]$. Let $x,y,z \in B$. The only cases we need to consider are $||\{x,y,z\}|| = 1$ and $||\{x,y,z\}|| = 2$. If $||\{x,y,z\}|| = 1$, i.e., $x = y = z$, then $\textit{set-g}(x,\textit{set-g}(y,z)) = \textit{set-g}(y,\textit{set-g}(x,z)) = \textit{set-g}(z,\textit{set-g}(x,y)) = \{x\}$ since $g$ is total on $B$ and self-contained. If $||\{x,y,z\}|| = 2$, assume without loss of generality that $x = z$ and $x \neq y$. It follows that $\textit{set-g}(y,\textit{set-g}(x,z)) = \textit{set-g}(y,x)$ since $g$ is total on $B$ and self-contained, and $\textit{set-g}(y,x) = \textit{set-g}(x,y)$ since $g$ is commutative. Observe that $\textit{set-g}(x,\textit{set-g}(y,z)) = \textit{set-g}(x,\textit{set-g}(y,x)) = \textit{set-g}(x,\textit{set-g}(x,y)) = \textit{set-g}(z,\textit{set-g}(x,y))$ since $x = z$ and $g$ is commutative. It remains to show that $\textit{set-g}(z,\textit{set-g}(x,y)) = \textit{set-g}(y,\textit{set-g}(x,z))$, i.e., in the current setting, showing that $\textit{set-g}(x,\textit{set-g}(x,y)) = \textit{set-g}(x,y)$. Note that $\textit{set-g}(x,x) = \{x\}$ and $\textit{set-g}(x,y) \in \{\{x\},\{y\},\{x,y\}\}$ since $g$ is total on $B$. If $\textit{set-g}(x,y) = \{x\}$ then $\textit{set-g}(x,\textit{set-g}(x,y)) = \textit{set-g}(x,y) = \{x\}$. If $\textit{set-g}(x,y) = \{y\}$ then $\textit{set-g}(x,\textit{set-g}(x,y)) = \textit{set-g}(x,y) = \{y\}$. If $\textit{set-g}(x,y) = \{x,y\}$ then $\textit{set-g}(x,\textit{set-g}(x,y)) = \textit{set-g}(x,y) = \{x,y\}$. We have shown that part 3 implies part 2. ❑

In particular note that, by Proposition 3.1, for any commutative self-contained function $g$ that is total on some set $B$, $g$ is associative on every one- or two-element set of strings from $B$, since item 3 of Proposition 3.1 is trivially satisfied when $||\{a,b,c\}|| < 3$.

Let $f$ be a (single- or multivalued) commutative self-contained function, $f : B \times B \to B$ for a finite set $B$. Consider the directed graph $G = (B,E)$, where $E = \{(x,y) \mid x,y \in B \land y \in \textit{set-f}(x,y)\}$. If $f$ is total on $B$ then $G$ is a complete digraph. If $f$ is single-valued and total on $B$ then $G$ is even an s-tournament. We will call $G$ the $f$-induced digraph on $B$. If $f$ is single-valued and total on $B$ then $G$ will be called the $f$-induced s-tournament on $B$. Note that the functions $f$ spoken of in Proposition 3.2 may be multivalued.

**Proposition 3.2** *Let $B$ be a finite set and $f$ be a commutative self-contained function that is total on $B$. $f$ is associative on $B$ if and only if $E = \{(x,y) \mid x,y \in B \land y \in \textit{set-f}(x,y)\}$, the edge set of the $f$-induced digraph, is a transitive relation.*

**Proof:** Let $B$ be a finite set and $f$ be a commutative self-contained function that is total on $B$.

We first prove the "only if" direction. So let $f$ be associative on $B$, i.e., for all $x,y,z \in B$, $\textit{set-f}(x,\textit{set-f}(y,z)) = \textit{set-f}(\textit{set-f}(x,y),z)$. Suppose that $E$ is not transitive. This means that there exits strings $a,b,c \in B$ such that $(a,b) \in E$ and $(b,c) \in E$ yet $(a,c) \notin E$. It follows that $a \neq b$ (since if $a = b$ then $(b,c) = (a,c)$), $b \neq c$ (since if $b = c$ then $(a,b) = (a,c)$), and $a \neq c$ (since if $a = c$ then $(a,a) = (a,c)$ and since $f$ is total on $B$ $(a,a) \in E$). So $a,b,c \in B$ are pairwise different strings such that $(a,b) \in E$, $(b,c) \in E$, and $(a,c) \notin E$. We will derive a contradiction by showing that $f$ is not associative, i.e., $\textit{set-f}(\textit{set-f}(a,b),c) \neq \textit{set-f}(a,\textit{set-f}(b,c))$. We do so by showing $c \in \textit{set-f}(\textit{set-f}(a,b),c)$ and $c \notin \textit{set-f}(a,\textit{set-f}(b,c))$. We have $b \in \textit{set-f}(a,b)$ and $c \in \textit{set-f}(b,c)$, and thus $c \in \textit{set-f}(\textit{set-f}(a,b),c)$. On the other hand, $c \notin \textit{set-f}(a,b)$ since $f$ is a self-contained function and $c \neq a$ and $c \neq b$. We also have $c \notin \textit{set-f}(a,c)$ since $(a,c) \notin E$. Since $\textit{set-f}(a,\textit{set-f}(b,c)) \subseteq \textit{set-f}(a,b) \cup \textit{set-f}(a,c)$ we have $c \notin \textit{set-f}(a,\textit{set-f}(b,c))$.

We now prove the "if" direction. So suppose that $E$ is transitive. We have to show that $f$ is associative on $B$, i.e., for any strings $a,b,c \in B$ we have to show $\textit{set-f}(a,\textit{set-f}(b,c)) = \textit{set-f}(\textit{set-f}(a,b),c)$. By Proposition 3.1 it suffices to show that, for all strings $a,b,c \in B$ such that $||\{a,b,c\}|| = 3$, $\textit{set-f}(a,\textit{set-f}(b,c)) = \textit{set-f}(b,\textit{set-f}(a,c)) = \textit{set-f}(c,\textit{set-f}(a,b))$. So suppose $a,b,c \in B$ are strings such that $||\{a,b,c\}|| = 3$.

We show $set\text{-}f(c, set\text{-}f(a,b)) \subseteq set\text{-}f(a, set\text{-}f(b,c))$. (The proof of the other inclusions follows via variable renaming and commutativity.) Consider $w \in set\text{-}f(c, set\text{-}f(a,b))$. If $w = a$ we have $a \in set\text{-}f(c, set\text{-}f(a,b))$ which implies, since $a \neq c$, $a \in set\text{-}f(a,b)$ and implies, since $a \notin \{b,c\}$, $a \in set\text{-}f(c,a) = set\text{-}f(a,c)$. Since $set\text{-}f(a,b) \cap set\text{-}f(a,c) \subseteq set\text{-}f(a, set\text{-}f(b,c))$ it follows that $a \in set\text{-}f(a, set\text{-}f(b,c))$. Similarly, if $w = b$ we have $b \in set\text{-}f(c, set\text{-}f(a,b))$ which in turn implies (since $b \neq c$) $b \in set\text{-}f(a,b)$ and implies (since $b \notin \{a,c\}$ and $b \in set\text{-}f(a,b)$) $b \in set\text{-}f(c,b) = set\text{-}f(b,c)$. It follows that $set\text{-}f(a,b) \subseteq set\text{-}f(a, set\text{-}f(b,c))$ and hence $b \in set\text{-}f(a, set\text{-}f(b,c))$. Finally suppose $w = c$ which yields $c \in set\text{-}f(c, set\text{-}f(a,b))$. It follows (since $c \notin \{a,b\}$) that either (a) $a \in set\text{-}f(a,b)$ and $c \in set\text{-}f(c,a) = set\text{-}f(a,c)$ or (b) $b \in set\text{-}f(a,b)$ and $c \in set\text{-}f(c,b) = set\text{-}f(b,c)$. If (a) holds we have $(b,a) \in E$ and $(a,c) \in E$. By the transitivity of $E$ it follows that also $(b,c) \in E$. By the definition of $E$ this implies $c \in set\text{-}f(b,c)$, and thus $c \in set\text{-}f(a, set\text{-}f(b,c))$. If (b) holds we have $(a,b) \in E$ and $(b,c) \in E$. Hence also $(a,c) \in E$ since $E$ is transitive. It follows that $c \in set\text{-}f(a,c)$, and thus $c \in set\text{-}f(a, set\text{-}f(b,c))$  ❑

Tournaments such that their edge set is a transitive relation are called transitive tournaments. Transitive tournaments implicitly impose a linear order on their nodes. We will use the following result (which can be viewed as a tournament-language expression of the fact that each finite linear order has a minimal and also a maximal element).

**Proposition 3.3** ([Moo68][3]) *The following statements are equivalent for s-tournaments $G = (V, E)$:*

1. *$G$ is a transitive s-tournament.*

2. *$G$ contains no directed cycles of length greater than 1.*

3. *Every induced (directed) subgraph $G' = (V', E')$, $V' \neq \emptyset$, of $G$ contains a (distance-one) source node, i.e., a node $s \in V'$ such that for all $u \in V'$, $(s,u) \in E'$.*

4. *Every induced (directed) subgraph $G' = (V', E')$, $V' \neq \emptyset$, of $G$ contains a (distance-one) target node, i.e., a node $t \in V'$ such that for all $u \in V'$, $(u,t) \in E'$.*

From Propositions 3.1, 3.2, and 3.3 we immediately obtain the following corollary.

**Corollary 3.4** *Let $f$ be a single-valued, commutative self-contained function that is total on $\{a,b,c\}$. $f$ is not associative on $\{a,b,c\}$ if and only if both $||\{a,b,c\}|| = 3$ and the $f$-induced s-tournament on $\{a,b,c\}$ is a 3-cycle (plus three self-loops).*

A statement similar to that of Proposition 3.3 can also be made for complete digraphs. But in contrast to the linear order implicitly given by a transitive s-tournament, a transitive complete digraph gives a linear order on a partition of its set of nodes, while the nodes of each part of the partition are in some sense pairwise indistinguishable.

Without giving all the definitions, we mention that in graph theory textbooks (see [BJG00], for instance) one can find the fact that the strongly connected components of a general digraph form a partial order. If we consider a complete digraph we immediately have that the strongly connected components form an order. The additional assumption of transitivity then yields that the strongly connected components are actually strong cliques. However, for reasons of self-containment we state and prove that graph-theoretic result below in a way that suits our purposes best.

**Proposition 3.5** *The following statements are equivalent for complete digraphs $G = (V, E)$:*

---
[3]Moon stated the result for tournaments, and so his statement lacks the "length greater than 1" condition in statement 2.

1. $E$ is a transitive relation.

2. For every $k \geq 1$, if $(a_1, a_2), (a_2, a_3), \ldots, (a_{k-1}, a_k), (a_k, a_1) \in E$ then $(a_i, a_j) \in E$ for all $1 \leq i, j \leq k$. In other words, if $a_1 \to a_2 \to \cdots \to a_{k-1} \to a_k \to a_1$ is a directed cycle in $G$, then the nodes $a_1, a_2, \ldots, a_k$ form a strong clique.

3. Every induced (directed) subgraph $G' = (V', E')$, $V' \neq \emptyset$, of $G$ contains a nonempty strong distance-one source-clique, i.e., an induced subgraph $S = (V_S, E_S)$, $V_S \neq \emptyset$, of $G'$ having the following properties:

    (a) $S$ is a maximal strong clique, i.e., for every $v \in V' - V_S$, the induced graph on the vertices $V_S \cup \{v\}$ is not a strong clique.

    (b) For all $u \in V_S$ and for all $v \in V' - V_S$, $(u, v) \in E'$ and $(v, u) \notin E'$.

4. Every induced (directed) subgraph $G' = (V', E')$, $V' \neq \emptyset$, of $G$ contains a nonempty strong distance-one target-clique, i.e., an induced subgraph $T = (V_T, E_T)$, $V_T \neq \emptyset$, of $G'$ having the following properties:

    (a) $T$ is a maximal strong clique, i.e., for every $v \in V' - V_T$, the induced graph on the vertices $V_T \cup \{v\}$ is not a strong clique.

    (b) For all $u \in V_T$ and for all $v \in V' - V_T$, $(u, v) \notin E'$ and $(v, u) \in E'$.

**Proof:** Let $G = (V, E)$ be a complete digraph. If $V = \emptyset$ all four conditions hold, so in this proof we henceforward assume $V \neq \emptyset$.

*Statement 1 implies statement 2.* Suppose that $E$ is transitive. Transitivity implies that for any vertices $u, v$ we have that the existence of a directed path from $u$ to $v$ implies $(u, v) \in E$. (This can easily be shown by induction on the length of the path connecting $u$ and $v$.) Now consider a cycle $a_1 \to a_2 \to \cdots \to a_{k-1} \to a_k \to a_1$. Because for any vertices $a_i, a_j$ there is a path from $a_i$ to $a_j$, we have $(a_i, a_j) \in E$. Hence the nodes $a_1, a_2, \ldots, a_k$ form a strong clique.

*Statement 2 implies both statement 3 and statement 4.* Suppose that statement 2 holds for a complete digraph $G = (V, E)$. Let $G' = (V', E')$, $V' \neq \emptyset$, be an induced subgraph of $G$. There are strong cliques in $G'$, because in a complete digraph every vertex forms a one-node strong clique. For the same reason, $G'$ has no zero-node maximal strong cliques. We show that maximal strong cliques are disjoint. If two maximal strong cliques $C$ and $C'$ have a vertex in common, then there is a cycle connecting all vertices in $C$ and $C'$. Then statement 2 implies that $C \cup C'$ forms a strong clique, contradicting the maximality of $C$ and $C'$. Also, for two maximal strong cliques $C$ and $C'$ all edges between them lead from $C$ to $C'$, or all of them lead from $C'$ to $C$, because otherwise there would be a cycle through all vertices of $C$ and $C'$, contradicting the maximality of $C$ and $C'$. Since every vertex $v \in V'$ is in a maximal strong clique, the maximal strong cliques partition $V'$. Define a relation $\preceq$ on the maximal strong cliques by $C \preceq C'$ if and only if there are $a \in C$ and $a' \in C'$ with $(a, a') \in E$. Note that any two maximal strong cliques are comparable with respect to this relation, since $G$ is a complete directed graph. Furthermore, the relation $\preceq$ is reflexive, antisymmetric, and transitive. Reflexivity and antisymmetry are obvious, transitivity can be seen as follows: Suppose $C \preceq C'$ and $C' \preceq C''$ holds for three maximal strong cliques $C$, $C'$, and $C''$ of $G'$. We have to show that then also $C \preceq C''$. If $C = C''$ then $C \preceq C''$ holds trivially since $\preceq$ is reflexive. If $C \neq C''$ then either $C \preceq C''$ or $C'' \preceq C$, but not both since $\preceq$ is antisymmetric. If $C \preceq C''$ we are done. If $C'' \preceq C$, then it indeed holds that there is a cycle through all vertices of $C \cup C' \cup C''$. So, by statement 2, $C \cup C' \cup C''$ forms a strong clique, contradicting the maximality of $C$, $C'$, and $C''$. Thus the transitivity of $\preceq$ is proven.

We have shown that the maximal strong cliques of $G'$ are linearly ordered by $\preceq$. Let $C_{\min}$ be the minimum element, and $C_{\max}$ be the maximum element in this order. ($C_{\min}$ and $C_{\max}$ do exist since $G'$ is a (finite) graph and thus contains only a finite number of maximal strong cliques.) Then $C_{\min}$ witnesses statement 3 and $C_{\max}$ witnesses statement 4.

*Statement 3 implies statement 1.* Let $a, b, c \in V$ and suppose $(a, b) \in E$ and $(b, c) \in E$. We will show that also $(a, c) \in E$. Note, if $a = b$ or $a = c$ or $b = c$ then $(a, c) \in E$ is trivially satisfied. So let $||\{a, b, c\}|| = 3$ and consider the induced subgraph consisting of the three nodes $a$, $b$, and $c$, call it $G_{[a,b,c]} = (\{a, b, c\}, E_{abc})$. According to statement 3 there is a subgraph $S = (V_S, E_S)$, $V_S \subseteq \{a, b, c\}$, satisfying (a) $S$ is a maximal strong clique and (b) for all $u \in V_S$ and for all $v \in \{a, b, c\} - V_S$, $(u, v) \in E_{abc}$ and $(v, u) \notin E_{abc}$. Hence, if $c \in V_S$ then $b \in V_S$, and if $b \in V_S$ then also $a \in V_S$. Since $V_S \neq \emptyset$, we can therefore conclude $a \in V_S$. It follows that $(a, c) \in E$.

Quite similarly to the proof of *statement 3 implies statement 1*, one can show *statement 4 implies statement 1*. ❑

Combining Proposition 3.2 and Proposition 3.5 we get the following corollary.

**Corollary 3.6** *Let $f$ be a (potentially multivalued) commutative self-contained function that is total on $\{a, b, c\}$. $f$ is not associative on $\{a, b, c\}$ if and only if both $||\{a, b, c\}|| = 3$ and the $f$-induced digraph on $\{a, b, c\}$ contains exactly one of the cycles $a \to b \to c \to a$ and $a \leftarrow b \leftarrow c \leftarrow a$.*

**Proof:** Let $f$ be a multivalued, commutative self-contained function that is total on a set $\{a, b, c\}$. Let $G = (\{a, b, c\}, E)$ be the $f$-induced digraph on $\{a, b, c\}$.

Suppose that $f$ is not associative on $\{a, b, c\}$. By the comment made immediately after the proof of Proposition 3.1 it follows that $||\{a, b, c\}|| = 3$. It follows from Proposition 3.2 that the edge set of the $f$-induced digraph is not a transitive relation and thus the $f$-induced digraph does not satisfy statement 2 of Proposition 3.5. It is not hard to see that this can only be the case if the $f$-induced digraph on $\{a, b, c\}$ contains exactly one of the cycles $a \to b \to c \to a$ and $a \leftarrow b \leftarrow c \leftarrow a$.

Now assume that $||\{a, b, c\}|| = 3$ and the $f$-induced digraph on $\{a, b, c\}$ contains exactly one of the cycles $a \to b \to c \to a$ and $a \leftarrow b \leftarrow c \leftarrow a$. By Proposition 3.5 we conclude that the edge set of the $f$-induced digraph is not transitive which in turn, by Proposition 3.2, implies that $f$ is not associative on $\{a, b, c\}$. ❑

Later in this paper, we will establish advice bounds in part by using the fact that associativity of a selector function (a restricted form of self-contained function) implies a linear structure on the underlying set.

## 4 Adding Commutativity to Associativity Is Free

Are all associatively P-selective sets commutatively, associatively P-selective? If they are, this will not only collapse two of our classes, but also will allow us to use results from the previous section in our study of associative selector functions. Unfortunately, Fact 2.3 (P-sel = C-P-sel) does *not* say that A-P-sel = CA-P-sel. The reason it does not guarantee this is that it is conceptually possible that the transformation denoted there by (∗) will, while gaining commutativity, destroy associativity. Theorem 4.1 and its proof establish that, for the case of $FP_t$ and $NPSV_t$-selector functions, this conceptual possibility is *not* a reality.

**Theorem 4.1**  *1.* A-P-sel = CA-P-sel.

*2.* A-NP-sel = CA-NP-sel.

**Proof:** As noted earlier, if $B$ is P-selective (NP-selective) via a P-selector (NP-selector) function $f$ then the function $f'(x,y) = f(\min(x,y), \max(x,y))$ is a commutative P-selector (NP-selector) for $B$. We show that the above transformation from $f$ to $f'$ preserves associativity when $f$ is a P-selector (NP-selector) function. Let $B \in$ A-P-sel ($B \in$ A-NP-sel) via an associative $FP_t$-selector ($NPSV_t$-selector) function $f$. It is clear that $f' \in FP_t$ ($f' \in NPSV_t$) and that $f'$ is commutative. As mentioned above, $f'$ is also a P-selector (NP-selector) for $B$. It remains to show that $f'$ is associative.

Suppose that $f'$ is not associative. According to Proposition 3.1 there exist three strings $a$, $b$, and $c$ such that $||\{a,b,c\}|| = 3$ and $f'(a, f'(b,c)) = f'(b, f'(a,c)) = f'(c, f'(a,b))$ does not hold. Without loss of generality assume $a <_{\text{lex}} b <_{\text{lex}} c$. It follows from Corollary 3.4 that the $f'$-induced s-tournament on $\{a,b,c\}$ is a 3-cycle (plus three self-loops). Suppose that the cycle is directed from $a$ to $b$, from $b$ to $c$, and back to $a$. This implies that $f'(a,b) = b$, $f'(b,c) = c$, and $f'(a,c) = a$. However, due to the definition of $f'$ this implies that $f(a,b) = f'(a,b) = b$, $f(b,c) = f'(b,c) = c$, and $f(a,c) = f'(a,c) = a$ which contradicts the associativity of $f$ since $f(a, f(b,c))) = a$ yet $f(f(a,b), c) = c$. Similarly, if the cycle is directed from $c$ to $b$, from $b$ to $a$, and back to $c$ then $f'(a,b) = a$, $f'(b,c) = b$, and $f'(a,c) = c$. According to the definition of $f'$ this implies $f(a,b) = f'(a,b) = a$, $f(b,c) = f'(b,c) = b$, and $f(a,c) = f'(a,c) = c$ which contradicts the associativity of $f$ since $f(a, f(b,c))) = a$ yet $f(f(a,b), c) = c$. This shows that $f'$ is associative. ❑

Similarly we have the following result.

**Theorem 4.2**   *1. $A_\ell$-P-sel = $A_\ell$C-P-sel.*

*2. $A_\ell$-NP-sel = $A_\ell$C-NP-sel.*

**Proof:** If a selector function $f$ is associative at each length, then we may invoke the entire proof of Theorem 4.1, except only seeking a counterexample (to the "associativity at each length" of $f'(x,y) = f(\min(x,y), \max(x,y))$) among triples of strings all of the same length. As in that proof, if such a counterexample exists (among some three strings of the same length) it forms a 3 cycle (plus three self-loops) in the associated directed graph. However, exactly as in the proof of Theorem 4.1, that contradicts the assumed associativity-at-each-length of $f$.[4]   ❑

So the transformation from a selector $f$ to a selector $f'$ via $f'(x,y) = f(\min(x,y), \max(x,y))$ not only provides commutativity but also preserves associativity if $f$ is total and single-valued. We mention that for this same case another standard transformation providing commutativity (namely, $f''(x,y) = \max(f(x,y), f(y,x))$) also has the associativity-preserving property.

However, the picture changes dramatically if one looks at multivalued selector functions $f$, for instance, $NPMV_t$ selector functions. Though for the second transformation it is not even clear what is meant by $\max(f(x,y), f(y,x))$ and whether that is again a function from $NPMV_t$, it can be shown that the first transformation does *not* preserve associativity.

**Proposition 4.3** *There are total associative multivalued selector functions $f$ such that the selector function $f'$, defined by $f'(x,y) = f(\min(x,y), \max(x,y))$, is not associative.*

---
[4]Note that in the above proof, though commutativity of $f'$ follows immediately from the definition of $f'$, showing that $f'$ is associative at each length relies on the fact that $f$ is total at each length. However, the following strengthening of Theorem 4.2 clearly follows from the proof of Theorem 4.2: For every FP-selector (respectively, NPSV-selector) $f$ for a set $B$ that is associative at each length, $f'$ is a a commutative FP-selector (respectively, NPSV-selector) for $B$ that is associative at each length $n$ at which $f$ is total.

However, we note in passing that one cannot strengthen Theorem 4.2 further to $A_\ell$-FP-sel = $A_\ell$C-FP-sel and $A_\ell$-NPSV-sel = $A_\ell$C-NPSV-sel by exploiting $f'$. A counterexample is the following: For strings $a, b, c$, $b <_{lex} c <_{lex} a$ all of the same length let $\textit{set-}f(a,b) = \textit{set-}f(b,a) = \textit{set-}f(a,c) = \textit{set-}f(c,b) = \emptyset$ and $\textit{set-}f(c,a) = \textit{set-}f(b,c) = \{c\}$. One can easily check that $f$ is associative on the three element set $\{a,b,c\}$. But $f'(a, f'(b,c)) = c$ yet $f'(f'(a,b), c) = \emptyset$. Similarly to the counterexample showing that $f'$ does not preserve associativity for multivalued functions (Proposition 4.3), one can extend the definition of $f$ to cover all strings from $\Sigma^*$.

**Proof:** Set $a = \epsilon$, $b = 0$, and $c = 1$. Consider the total multivalued selector function $f$ that has the following values on $a$, $b$, and $c$ and thus induces the following values of $f'$ on $a$, $b$, and $c$:

$$\begin{array}{ll}
\text{set-}f(a,b) = \{a\} & \text{set-}f(b,a) = \{b\} \\
\text{set-}f(a,c) = \{a,c\} & \text{set-}f(c,a) = \{c\} \\
\text{set-}f(b,c) = \{b,c\} & \text{set-}f(c,b) = \{c\}
\end{array} \quad \begin{array}{l}
\text{set-}f'(a,b) = \text{set-}f'(b,a) = \text{set-}f(a,b) = \{a\} \\
\text{set-}f'(a,c) = \text{set-}f'(c,a) = \text{set-}f(a,c) = \{a,c\} \\
\text{set-}f'(b,c) = \text{set-}f'(c,b) = \text{set-}f(b,c) = \{b,c\}
\end{array}$$

Furthermore, for each $x \in \{a,b,c\}$ and each $y \in \Sigma^* - \{a,b,c\}$ let $\text{set-}f(x,y) = \text{set-}f(y,x) = \{y\}$. And for $y, z \in \Sigma^* - \{a,b,c\}$ let $\text{set-}f(y,z) = \text{set-}f(z,y) = \{\max(y,z)\}$. Note that if $\{e, e'\} \not\subseteq \{a,b,c\}$ then we have $\text{set-}f(e,e') = \text{set-}f'(e,e')$.

It is not hard to verify that $f$ is indeed associative, due to the way any non-$\{a,b,c\}$ argument "trumps" any $\{a,b,c\}$ argument, and regarding all-$\{a,b,c\}$ arguments we have

$$\begin{aligned}
\text{set-}f(a, \text{set-}f(b,c)) &= \text{set-}f(\text{set-}f(a,b), c) = \{a,c\}, \\
\text{set-}f(a, \text{set-}f(c,b)) &= \text{set-}f(\text{set-}f(a,c), b) = \{a,c\}, \\
\text{set-}f(b, \text{set-}f(a,c)) &= \text{set-}f(\text{set-}f(b,a), c) = \{b,c\}, \\
\text{set-}f(b, \text{set-}f(c,a)) &= \text{set-}f(\text{set-}f(b,c), a) = \{b,c\}, \\
\text{set-}f(c, \text{set-}f(a,b)) &= \text{set-}f(\text{set-}f(c,a), b) = \{c\}, \text{ and} \\
\text{set-}f(c, \text{set-}f(b,a)) &= \text{set-}f(\text{set-}f(c,b), a) = \{c\}.
\end{aligned}$$

But $f'$ is not associative since

$$\text{set-}f'(a, \text{set-}f'(c,b)) = \{a,c\} \quad \text{and} \quad \text{set-}f'(\text{set-}f'(a,c), b) = \{a,b,c\}. \qquad \square$$

Observe that we have shown that by the mentioned transformation from $f$ to $f'$ that gains in commutativity can lose associativity. However, for total multivalued functions there is another transformation that does preserve associativity while adding commutativity, namely $\text{set-}\widehat{f}(x,y) = \text{set-}f(x,y) \cup \text{set-}f(y,x)$, for all $x$ and $y$. Theorem 4.4 establishes this.

**Theorem 4.4** A-NPMV$_t$-sel = CA-NPMV$_t$-sel.

**Proof:** It suffices to show A-NPMV$_t$-sel $\subseteq$ CA-NPMV$_t$-sel. Let $B \in$ A-NPMV$_t$-sel and let $f \in$ NPMV$_t$ be an associative selector for $B$. We will now show that $\widehat{f}$ is a commutative and associative NPMV$_t$-selector for $B$.

It is not hard to see that $\widehat{f}$ is commutative, $\widehat{f} \in$ NPMV$_t$, and that indeed $\widehat{f}$ is an NPMV$_t$-selector for $B$. It remains to show that $\widehat{f}$ is associative. By Proposition 3.1 it suffices to show that $\widehat{f}$ is associative on every set $\{a,b,c\}$ such that $\|\{a,b,c\}\| = 3$. By Corollary 3.6 it suffices to show that the $\widehat{f}$-induced graph on $\{a,b,c\}$, call it $G_{[a,b,c]} = (\{a,b,c\}, E_{abc})$, contains both the cycles $a \to b \to c \to a$ and $a \leftarrow b \leftarrow c \leftarrow a$ if it contains at least one of them.

Suppose $G_{[a,b,c]}$ contains $a \to b \to c \to a$ (the case where $G_{[a,b,c]}$ contains $a \leftarrow b \leftarrow c \leftarrow a$ is analogous under renaming so we do not separately cover it). We have to show that then also the other cycle is in $G_{[a,b,c]}$, i.e., $(a,c)$, $(c,b)$, and $(b,a)$ are in $E_{abc}$. For symmetry reasons, it suffices to show that $(a,c) \in E_{abc}$ or, equivalently, $c \in \text{set-}\widehat{f}(a,c)$. Because $(a,b) \in E_{abc}$, we have $b \in \text{set-}\widehat{f}(a,b)$ and thus $b \in \text{set-}f(a,b)$ or $b \in \text{set-}f(b,a)$. Similarly, because $(b,c) \in E_{abc}$, we have $c \in \text{set-}\widehat{f}(b,c)$ and thus $c \in \text{set-}f(b,c)$ or $b \in \text{set-}f(c,b)$. We distinguish four cases.

**Case 1** [$b \in \text{set-}f(a,b)$ **and** $c \in \text{set-}f(b,c)$]:

This implies $c \in \text{set-}f(\text{set-}f(a,b), c)$. Since $f$ is associative, $c \in \text{set-}f(a, \text{set-}f(b,c))$. Since $f$ is self-contained and thus $\text{set-}f(a, \text{set-}f(b,c)) \subseteq \text{set-}f(a,b) \cup \text{set-}f(a,c)$ we conclude $c \in \text{set-}f(b,c)$ and $c \in \text{set-}f(a,c)$. Therefore $c \in \text{set-}\widehat{f}(a,c)$.

**Case 2** $[b \in \text{set-}f(a,b) \text{ and } c \in \text{set-}f(c,b)]$:

This implies $c \in \text{set-}f(c, \text{set-}f(a,b))$. Since $f$ is associative, $c \in \text{set-}f(\text{set-}f(c,a),b)$. It follows that $c \in \text{set-}f(c,a)$ since $f$ is self-contained. Therefore $c \in \text{set-}\widehat{f}(a,c)$.

**Case 3** $[b \in \text{set-}f(b,a) \text{ and } c \in \text{set-}f(b,c)]$:

This implies $c \in \text{set-}f(\text{set-}f(b,a),c)$. Since $f$ is associative, $c \in \text{set-}f(b,\text{set-}f(a,c))$. Again, it follows that $c \in \text{set-}f(a,c)$ since $f$ is self-contained. Therefore $c \in \text{set-}\widehat{f}(a,c)$.

**Case 4** $[b \in \text{set-}f(b,a) \text{ and } c \in \text{set-}f(c,b)]$:

This implies $c \in \text{set-}f(c,\text{set-}f(b,a))$. Since $f$ is associative, $c \in \text{set-}f(\text{set-}f(c,b),a)$. It follows that $c \in \text{set-}f(c,b)$ since $f$ is self-contained and thus $c \in \text{set-}f(c,a)$. Therefore $c \in \text{set-}\widehat{f}(a,c)$.

It follows that $(a,c) \in E_{abc}$. This completes the proof that $\widehat{f}$ is associative. ❑

We also have the following.

**Theorem 4.5** $A_\ell\text{-NPMV}_t\text{-sel} = A_\ell C\text{-NPMV}_t\text{-sel}$.

The proof of this theorem proceeds in analogy to the proof of Theorem 4.4 except focusing on strings of equal length.[5]

Finally we turn to partial selector functions. The definition of selector functions allows a partial selector function for a set $B$ to be undefined only when both inputs are not contained in $B$. However, the additional requirement of associativity forces any selector function for a nonempty set to be outright total.

**Proposition 4.6**   1. *If a (single- or multivalued) selector function $f$ for a set $B \neq \emptyset$ is associative then $f$ is a total function.*

2. *If a (single- or multivalued) selector function $f$ for a set $B$ is associative at each length then $f(x,y)$ is defined for all $x$ and $y$ such that both $|x| = |y|$ and $B^{=|x|} \neq \emptyset$.*

**Proof:** We first prove part 1 of the proposition. Let $f$ be an associative selector function for a set $B \neq \emptyset$. For any inputs $x$ and $y$ such that $\{x,y\} \cap B \neq \emptyset$, $f(x,y)$ is not undefined (equivalently, $\text{set-}f(x,y) \neq \emptyset$), due to the definition of selectivity. Assume $x,y \notin B$ and $f(x,y)$ is undefined, i.e., $\text{set-}f(x,y) = \emptyset$. Consider a third string $z \in B$. Such a string exists since $B \neq \emptyset$. Note that $\text{set-}f(x,z) = \{z\}$ and $\text{set-}f(y,z) = \{z\}$ according to the definition of a selector function. So $\text{set-}f(x,\text{set-}f(y,z)) = \{z\}$ yet (since $\text{set-}f(x,y) = \emptyset$) $\text{set-}f(\text{set-}f(x,y),z) = \emptyset$. This contradicts the associativity of $f$.

The proof of part 2 is very similar to the proof of the previous part. ❑

**Corollary 4.7**   1. A-FP-sel = A-FP$_t$-sel = CA-FP$_t$-sel. *(That is,* A-FP-sel = A-P-sel = CA-P-sel.*)*

---

[5] Comments similar to those of Footnote 4 apply here. Namely, the following strengthening of Theorem 4.5 clearly follows from the proof of Theorem 4.5: For every NPMV-selector $f$ for a set $B$ that is associative at each length, $\widehat{f}$ is a commutative NPMV-selector for $B$ that is associative at each length $n$ at which $f$ is total.

However, Theorem 4.5 can not be strengthened to $A_\ell$-NPMV-sel = $A_\ell C$-NPMV-sel using $\widehat{f}$. A counterexample is the following: For strings $a,b,c$ all of the same length let $\text{set-}f(a,b) = \text{set-}f(b,a) = \text{set-}f(a,c) = \text{set-}f(c,b) = \emptyset$ and $\text{set-}f(c,a) = \text{set-}f(b,c) = \{c\}$. One can easily check that $f$ is associative on the three element set $\{a,b,c\}$. But $\text{set-}\widehat{f}(a,\text{set-}\widehat{f}(b,c)) = \{c\}$ yet $\text{set-}\widehat{f}(\text{set-}\widehat{f}(a,b),c) = \emptyset$. Similarly to the counterexample showing that $f'$ does not preserve associativity for multivalued functions (Proposition 4.3), one can extend the definition of $f$ to cover all strings from $\Sigma^*$.

2. *Every set $B \in A_\ell$-FP-sel has a commutative FP-selector that is total at each length $n$ for which $B^{=n} \neq \emptyset$ and associative at each length $n$ for which $B^{=n} \neq \emptyset$.*

3. A-NPSV-sel = A-NPSV$_t$-sel = CA-NPSV$_t$-sel. *(That is* A-NPSV-sel = A-NP-sel = CA-NP-sel.*)*

4. *Every set $B \in A_\ell$-NPSV-sel has a commutative NPSV-selector that is total at each length $n$ for which $B^{=n} \neq \emptyset$ and associative at each length $n$ for which $B^{=n} \neq \emptyset$.*

5. A-NPMV-sel = A-NPMV$_t$-sel = CA-NPMV$_t$-sel.

6. *Every set $B \in A_\ell$-NPMV-sel has a commutative NPMV-selector that is total at each length $n$ for which $B^{=n} \neq \emptyset$ and associative at each length $n$ for which $B^{=n} \neq \emptyset$.*

Parts 1 and 3 follow directly from Proposition 4.6 and Theorem 4.1. Note that $\emptyset$ is trivially contained in CA-P-sel. Parts 2 and 4 follow implicitly from Proposition 4.6 and the proof of Theorem 4.2 (note footnote 4). Part 5 follows directly from Proposition 4.6 and Theorem 4.4. Part 6 again follows implicitly from Proposition 4.6 and the proof of Theorem 4.5 (note footnote 5).

## 5 Associativity Drops Advice Complexity to Linear

Ko showed the following result.

**Theorem 5.1 ([Ko83])** P-sel $\subseteq$ P/$\mathcal{O}(n^2)$.

That is, deterministic advice interpreters given quadratic advice accept the P-selective sets. It is also known that nondeterministic advice interpreters with linear advice accept the P-selective sets, and this is optimal.

**Theorem 5.2 ([HT96], see also [HNP98])** P-sel $\subseteq$ NP/$n+1$ *but* P-sel $\not\subseteq$ NP/$n$.

It is natural to wonder whether the "P" advice interpretation of Theorem 5.1 can be unified with the linear advice amount of Theorem 5.2 to obtain the very strong claim: P-sel $\subseteq$ P/$\mathcal{O}(n)$. Currently, no proof of this claim is known, people doing research on selectivity generally doubt that it holds, and there is some very recent relativized evidence in harmony with that doubt [Tha03]. However, proving the claim false seems unlikely in the immediate future, since by Theorem 5.2 any such proof would implicitly prove P $\neq$ NP.

Nonetheless, the following result shows that all associatively P-selective sets *are* accepted by deterministic advice interpreters with linear advice.

**Theorem 5.3** $A_\ell$-FP-sel $\subsetneq$ P/$n+1$.

(P/1) $-$ NPMV-sel $\neq \emptyset$ (see the comments in the proof below of Theorem 5.3 regarding why this holds). In light of this inequality, Theorem 5.3 supports the following corollary.

**Corollary 5.4** *1. A-P-sel $\subsetneq$ P/$n+1$.*

*2. $A_\ell$-P-sel $\subsetneq$ P/$n+1$.*

**Proof of Theorem 5.3:** Let $B \in A_\ell$-FP-sel. Let $f$ be a commutative FP-selector for $B$ that (a) is total at each length $n$ such that $B^{=n} \neq \emptyset$ and that (b) is associative at each length $n$ such that $B^{=n} \neq \emptyset$. By Corollary 4.7 we know that such a commutative FP-selector for $B$ must exist.

Let $n \in \mathbb{N}$ such that $B^{=n} \neq \emptyset$. Consider the directed graph $G_n = (B^{=n}, E_n)$, where $E_n = \{(x, y) \mid x, y \in B^{=n} \wedge f(x, y) = y\}$. $G_n$ is a nonempty s-tournament.

Since $f$ is a commutative selector function that is total and associative on $B^{=n}$ it follows from Proposition 3.2 that $E_n$ is a transitive relation. So $G_n$ is a transitive s-tournament, which by Proposition 3.3 implies that $G_n$ contains a source node $s_n$. Note that $x \in B^{=n}$ implies $f(x, s_n) = x$ and $x \in \Sigma^n - B^{=n}$ implies $f(x, s_n) = s_n$. In other words, for all $n \in \mathbb{N}$, if $B^{=n} \neq \emptyset$ then there exists a string $s_n \in B^{=n}$ such that for all $x \in \Sigma^n$,

$$x \in B \iff f(x, s_n) = x.$$

Define, for all $n \in \mathbb{N}$,

$$h(1^n) = \begin{cases} 1s_n & \text{if } B^{=n} \neq \emptyset \\ 0^{n+1} & \text{otherwise.} \end{cases}$$

Note that $h$ on any input $1^n$ outputs a string of length exactly $n+1$. Let $L = \{\langle x, 1y \rangle \mid |x| = |y| \wedge f(x, y) = x\}$. Clearly, $L \in \text{P}$. Due to the above construction we also have, for all $x \in \Sigma^*$,

$$x \in B \iff \langle x, h(1^{|x|}) \rangle \in L.$$

This shows that $B \in \text{P}/n+1$.

It remains to show that $\text{P}/n+1 - A_\ell$-FP-sel $\neq \emptyset$. In fact, it clearly holds even that $(\text{P}/1) - \text{NPMV-sel} \neq \emptyset$. The proof of this claim consists of a diagonalization against all self-contained NPMV functions. Let $f_1, f_2, \ldots$ be an enumeration (not necessarily effective, though in fact there do exist such effective enumerations) of all self-contained NPMV functions. Set

$$\begin{aligned} B &= \{0^{2i-1} \mid i \in \mathbb{N} \wedge \text{set-}f_i(0^{2i-2}, 0^{2i-1}) = \{0^{2i-2}\}\} \cup \\ &\quad \{0^{2i-2} \mid i \in \mathbb{N} \wedge \text{set-}f_i(0^{2i-2}, 0^{2i-1}) \neq \{0^{2i-2}\}\} . \end{aligned}$$

$B$ is not NPMV-selective, yet $B \in \text{P}/1$. ❑

Note that for the correctness of the proof of Theorem 5.3 it actually suffices for $f$ to be a partial FP-selector that is weakly associative at each length. $f$ being a selector for $B$ ensures that $G_n$ is an s-tournament since partial selectors for $B$ can be undefined only if both inputs are not in $B$. The weak associativity at each length of $f$ yields that $E_n$ is transitive. And finally, the above-defined set $L$ remains a P set even under those weaker assumptions.

We note also that, more generally than Theorem 5.3, if a set $A$ and a (not necessarily commutative) P-selector $f$ for it have the property that at each nonempty length $m$ there is some string $y \in A^{=m}$ at that length—or even at some length linearly related to that length—such that $f(y, z) = z$ or $f(z, y) = z$ for each $z \in A^{=m}$, then $A$ is in $\text{P}/\mathcal{O}(n)$. However, among ways of ensuring that this condition is met, we feel that associativity is a particularly natural and well-structured property to study.

The number of advice bits in Theorem 5.3 and Corollary 5.4, $n+1$, is optimal.

**Theorem 5.5** A-P-sel $\nsubseteq \text{P}/n$.

**Proof:** Hemaspaandra and Torenvliet [HT96] construct a P-selective set, call it $B$, such that $B \nsubseteq \text{P}/n$. We will show that their set is A-selective. (We will not repeat their proof that $B \nsubseteq \text{P}/n$, since they already proved that.) Hemaspaandra and Torenvliet do not completely specify the P-selector function for $B$, though they make it clear that $B$ is P-selective. In fact, it is easy to see that there are ways of completing the specification of their P-selector function in such a way that it is not associative. However, our task is merely to prove that there is at least one way to complete their P-selector function so as to make it associative. We now do so.

The set $B$ constructed by Hemaspaandra and Torenvliet satisfies the following properties:

1. $B \in \text{DTIME}[2^{2^n}]$,

2. For all strings $x$, if $x \in B$ then $|x| \in \mathcal{L}$, where $\ell_0 = 2$, $\ell_{i+1} = 2^{2^{2^{\ell_i}}}$ for all $i \geq 0$, and $\mathcal{L} = \{\ell_0, \ell_1, \ell_2, \ldots\}$.

3. For all strings $x$ and $y$ such that $|x| = |y|$, if $x \leq_{\text{lex}} y$ then $\chi_B(x) \leq \chi_B(y)$, where $\chi_B$ denotes the characteristic function of $B$.

Let $f$ be the following function.

$$f(x,y) = \begin{cases} x & \text{if } |x| \in \mathcal{L} \text{ and } |y| \notin \mathcal{L} \\ y & \text{if } |x| \notin \mathcal{L} \text{ and } |y| \in \mathcal{L} \\ x & \text{if } |x|, |y| \in \mathcal{L}, |x| < |y|, \text{ and } x \in B \\ y & \text{if } |x|, |y| \in \mathcal{L}, |x| < |y|, \text{ and } x \notin B \\ y & \text{if } |x|, |y| \in \mathcal{L}, |x| > |y|, \text{ and } y \in B \\ x & \text{if } |x|, |y| \in \mathcal{L}, |x| > |y|, \text{ and } y \notin B \\ \max(x,y) & \text{otherwise.} \end{cases}$$

Note that the "otherwise" case is reached exactly when

$$(|x| \notin \mathcal{L} \text{ and } |y| \notin \mathcal{L}) \text{ or } |x| = |y|.$$

In light of the above properties of $B$ it is not hard to verify that $f$ is polynomial-time computable. The key thing to note to see this is that when $|x|, |y| \in \mathcal{L}$ and $|x| \neq |y|$, then in that case the huge gaps in $\mathcal{L}$ and the fact that $B \in \text{DTIME}[2^{2^n}]$ allow us to brute-force test $\min(x,y) \in B$ in time polynomial in $|x| + |y|$. Also, (the polynomial-time computable function) $f$ is clearly a P-selector for $B$.

What remains to show is that $f$ is associative. From its definition, $f$ is clearly commutative. So by Proposition 3.1, it suffices to show that for all $x, y, z \in \Sigma^*$, $||\{x, y, z\}|| = 3$, it holds that $f(x, f(y,z)) = f(y, f(x,z)) = f(z, f(x,y))$. So let $a$, $b$, and $c$ be three strings such that $||\{a, b, c\}|| = 3$. Assume without loss of generality that $a <_{\text{lex}} b <_{\text{lex}} c$. Clearly it will also hold that $|a| \leq |b| \leq |c|$.

**Case 1** $[|a|, |b|, |c| \notin \mathcal{L}]$:
Applying the definition of $f$, we see that $f(a, f(b,c)) = f(b, f(a,c)) = f(c, f(a,b)) = \max(a, b, c)$.

**Case 2** [**The length of exactly one string from $\{a, b, c\}$ is in $\mathcal{L}$**]:
Let $w \in \{a, b, c\}$ be the unique string among $a$, $b$, and $c$ such that $|w| \in \mathcal{L}$. Let $v_1$ and $v_2$ be the two strings in $\{a, b, c\}$ such that $|v_1|, |v_2| \notin \mathcal{L}$. So $f(w, v_1) = w$, $f(w, v_2) = w$, and consequently, since $f$ is a self-contained function $f(w, f(v_1, v_2)) = f(v_1, f(w, v_2)) = f(v_2, f(w, v_1)) = w$. It follows that $f(a, f(b,c)) = f(b, f(a,c)) = f(c, f(a,b)) = w$.

**Case 3 [The length of exactly two strings from $\{a,b,c\}$ is in $\mathcal{L}$]:**
Let $v \in \{a,b,c\}$ be the unique string among $a$, $b$, and $c$ such that $|v| \notin \mathcal{L}$. Let $w_1$ and $w_2$ be the two strings in $\{a,b,c\}$ such that $|w_1|, |w_2| \in \mathcal{L}$. So $f(v, w_1) = w_1$, $f(v, w_2) = w_2$, and consequently $f(v, f(w_1, w_2)) = f(w_1, f(v, w_2)) = f(w_2, f(v, w_1)) = f(w_1, w_2)$. It follows that $f(a, f(b,c)) = f(b, f(a,c)) = f(c, f(a,b)) = f(w_1, w_2)$.

**Case 4 [$|a|, |b|, |c| \in \mathcal{L}$]:**

**Case 4.1 [$\|\{|a|, |b|, |c|\}\| = 1$]:**
So $|a| = |b| = |c|$ and thus, by the definition of $f$, we have $f(a, f(b,c)) = f(b, f(a,c)) = f(c, f(a,b)) = \max(a,b,c)$.

**Case 4.2 [$\|\{|a|, |b|, |c|\}\| = 2$]:**
So, as $|a| \le |b| \le |c|$, it holds that $|a| = |b|$ or $|b| = |c|$, yet not all three strings have the same length.

**Case 4.2.1 [$|a| = |b|$ and $|b| < |c|$]:**
Note that $f(a,b) = \max(a,b) = b$,

$$f(a,c) = \begin{cases} a & \text{if } a \in B \\ c & \text{otherwise,} \end{cases} \quad \text{and} \quad f(b,c) = \begin{cases} b & \text{if } b \in B \\ c & \text{otherwise.} \end{cases}$$

Recall that $a <_{\text{lex}} b <_{\text{lex}} c$. Due to the third property of $B$ quoted earlier in this proof, we thus have $a \in B \implies b \in B$ or, equivalently, $b \notin B \implies a \notin B$. So

$$f(a, f(b,c)) = f(b, (f(a,c)) = f(c, f(a,b)) = \begin{cases} b & \text{if } b \in B \\ c & \text{otherwise.} \end{cases}$$

**Case 4.2.2 [$|a| < |b|$ and $|b| = |c|$]:**
Recall that $a <_{\text{lex}} b <_{\text{lex}} c$. So $f(b,c) = \max(b,c) = c$,

$$f(a,b) = \begin{cases} a & \text{if } a \in B \\ b & \text{otherwise,} \end{cases} \quad \text{and} \quad f(a,c) = \begin{cases} a & \text{if } a \in B \\ c & \text{otherwise.} \end{cases}$$

It follows that

$$f(a, f(b,c)) = f(b, (f(a,c)) = f(c, f(a,b)) = \begin{cases} a & \text{if } a \in B \\ c & \text{otherwise.} \end{cases}$$

**Case 4.3 [$\|\{|a|, |b|, |c|\}\| = 3$]:**
So we have $|a|, |b|, |c| \in \mathcal{L}$ and $|a| < |b| < |c|$. Thus

$$f(a,b) = \begin{cases} a & \text{if } a \in B \\ b & \text{otherwise,} \end{cases} \quad f(a,c) = \begin{cases} a & \text{if } a \in B \\ c & \text{otherwise,} \end{cases}$$

$$\text{and} \quad f(b,c) = \begin{cases} b & \text{if } b \in B \\ c & \text{otherwise.} \end{cases}$$

It follows that

$$f(a, f(b,c)) = f(b, (f(a,c)) = f(c, f(a,b)) = \begin{cases} a & \text{if } a \in B \\ b & \text{if } a \notin B \text{ and } b \in B \\ c & \text{otherwise.} \end{cases}$$

This completes the proof of the claim that $f$ is associative. ❑

**Corollary 5.6**  *1.* $A_\ell$-P-sel $\not\subseteq$ P/$n$.

*2.* $A_\ell$-FP-sel $\not\subseteq$ P/$n$.

In fact, for any recursive function $g$, A-P-sel $\not\subseteq$ DTIME$[g(n)]/n$. The reason for this is that we may take the P-selective set that Hemaspaandra and Torenvliet [HT96] prove is not in DTIME$[g(n)]/n$ and, just as in the proof of Theorem 5.5, may complete its P-selector function in a manner that achieves associativity. It follows, from this observation, that the linear advice bound for associatively NP-selective sets and also the linear advice bound for associatively NPMV-selective sets that will be proven as Theorem 5.9 and Corollary 5.10 are optimal.

**Corollary 5.7**  *1.* A-NP-sel $\not\subseteq$ (NP $\cap$ coNP)/$n$.

*2.* $A_\ell$-NP-sel $\not\subseteq$ (NP $\cap$ coNP)/$n$.

*3.* $A_\ell$-NPSV-sel $\not\subseteq$ (NP $\cap$ coNP)/$n$.

*4.* A-NPMV$_t$-sel $\not\subseteq$ NP/$n$.

*5.* $A_\ell$-NPMV$_t$-sel $\not\subseteq$ NP/$n$.

*6.* $A_\ell$-NPMV-sel $\not\subseteq$ NP/$n$.

So we have established an $n$-bit lower bound for the advice complexity of (associatively) NP- and NPMV-selective sets. The following upper bounds on the amount of advice needed to (nondeterministically) accept NP- and NPMV-selective sets are known.

**Theorem 5.8**  *1.* [HHN$^+$95] NP-sel $\subseteq$ (NP $\cap$ coNP)/poly.

*2.* [HNOS96b] NPSV-sel $\subseteq$ NP/poly $\cap$ coNP/poly *and* NPSV-sel $\cap$ NP $\subseteq$ (NP $\cap$ coNP)/poly.

*3.* [HNOS96b] NPMV$_t$-sel $\subseteq$ NP/poly $\cap$ coNP/poly.

*4.* [HNOS96b] NPMV-sel $\subseteq$ NP/poly.

We note that the results we have stated that involve complexity classes relativize. In particular, one can relativize them by any particular NP $\cap$ coNP set, and so regarding associativity and the NP-selective sets (equivalently the NPSV$_t$-selective sets and equivalently the FP$^{NP \cap coNP}$-selective sets, see [HNOS96b,HHN$^+$95] for definitions and discussion), it follows from our results that all length-associatively NP-selective sets are in (NP $\cap$ coNP)/$n+1$.

**Theorem 5.9**  *1.* $A_\ell$-NP-sel $\subsetneq$ (NP $\cap$ coNP)/$n+1$.

*2.* $A_\ell$-NPMV-sel $\subsetneq$ NP/$n+1 \cap$ coNP/$n+1$.

**Proof:** Regarding the strictness ($\subsetneq$ rather than $\subseteq$) of both parts, this follows from the easy fact (noted in the proof of Theorem 5.3) that (P/1) $-$ NPMV-sel $\neq \emptyset$. The containment claims of both parts remain to be proven.

We now prove the containment part of item 1 of the theorem, i.e., that $A_\ell$-NP-sel $\subseteq$ (NP $\cap$ coNP)/$n+1$. Let $B \in A_\ell$-NP-sel. It follows from Theorem 4.2 that there exists a commutative NPSV$_t$-selector $f$ for $B$ that is associative at each length.

One can now invoke the entire containment part of the proof of Theorem 5.3, i.e., the proof for the claim $A_\ell$-P-sel $\subseteq$ P/$n+1$ (this is actually that proof weakened to the case described by part 2 of Corollary 5.4), with the obvious change from P-selector to NPSV$_t$-selector and thus resulting in the fact that the set $L$ defined in the proof of Theorem 5.3 is now in NP $\cap$ coNP. To see the latter simply note that $L = \{\langle x, 1y\rangle \mid |x| = |y| \wedge f(x,y) = x\}$ can also be written as $L = \{\langle x, 1y\rangle \mid |x| = |y| \wedge (x = y \vee f(x,y) \neq y)\}$ since $f$ is a total NPSV-selector.

We now show the containment part of item 2, i.e., that $A_\ell$-NPMV-sel $\subseteq$ NP/$n+1 \cap$ coNP/$n+1$. Let $B \in A_\ell$-NPMV-sel. It follows from Corollary 4.7 that there exists a commutative NPMV-selector $f$ for $B$ that is total at each length at which $B$ is nonempty, and that is associative at each length at which $B$ is nonempty.

Let $n \in \mathbb{N}$ be such that $B^{=n} \neq \emptyset$. Consider the directed graph $G_n = (B^{=n}, E_n)$, where $E_n = \{(x,y) \mid x, y \in B^{=n} \wedge y \in \text{set-}f(x,y)\}$. By the definition of NPMV-selectivity (see Definition 2.1), $f$ must be defined for all $x, y \in B^{=n}$. So $G_n$ is a nonempty complete digraph. In fact, note even that, from Proposition 4.6, $f$ is defined for all $x, y \in \Sigma^n$. Since $f$ is a commutative selector function that is total and associative on $B^{=n}$, it follows from Proposition 3.2 that $E_n$ is a transitive relation. So it follows from Proposition 3.5 that $G_n$ contains a nonempty strong distance-one source-clique, i.e., an induced subgraph $S = (V_S, E_S)$, $V_S \neq \emptyset$, having the following properties:

1. $S$ is a maximal strong clique.

2. For all $u \in V_S$ and for all $v \in B^{=n} - V_S$, $(u,v) \in E_n$ and $(v,u) \notin E_n$.

Let $u_n$ denote the lexicographically smallest string in $V_S$. (This is just for specificity. Any string from $V_S$ would function the same way.) Note that $x \in B^{=n}$ implies $x \in \text{set-}f(u_n, x)$ whereas $x \in \Sigma^n - B^{=n}$ implies $x \notin \text{set-}f(u_n, x)$. Hence, for all $n \in \mathbb{N}$ such that $B^{=n} \neq \emptyset$ there exists a string $u_n \in B^{=n}$ such that for all $x \in \Sigma^n$,

$$x \in B \iff x \in \text{set-}f(u_n, x).$$

Define, for all $n \in \mathbb{N}$,

$$h(1^n) = \begin{cases} 1u_n & \text{if } B^{=n} \neq \emptyset \\ 0^{n+1} & \text{otherwise.} \end{cases}$$

Note that $h$ on any input $1^n$ outputs a string of length exactly $n + 1$. Let $L = \{\langle x, 1y\rangle \mid |x| = |y| \wedge x \in \text{set-}f(y,x)\}$. Clearly, $L \in$ NP. Due to the above construction we also have, for all $x \in \Sigma^*$,

$$x \in B \iff \langle x, h(1^{|x|})\rangle \in L.$$

This shows that $B \in$ NP/$n+1$.

To see that $B \in$ coNP/$n+1$ we have to repeat the argument on $\Sigma^n - B^{=n}$. Let $n \in \mathbb{N}$ such that both $B^{=n} \neq \emptyset$ and $\Sigma^n - B^{=n} \neq \emptyset$. (If $B^{=n} = \emptyset$ or $\Sigma^n - B^{=n} = \emptyset$, then we at this length easily have a coNP/$n+1$ algorithm for that case, and will signal such cases explicitly in the advice function that we will soon build.) Consider the directed graph $G'_n = (\Sigma^n - B^{=n}, E'_n)$, where $E'_n = \{(x,y) \mid x, y \in \Sigma^n - B^{=n} \wedge y \in \text{set-}f(x,y)\}$. $G'_n$ is a nonempty complete digraph. ($G'_n$ is nonempty since $\Sigma^n - B^{=n} \neq \emptyset$ and $G'_n$ is a complete digraph since $f$ is total at length $n$ due to $B^{=n} \neq \emptyset$ and Proposition 4.6.) Since $f$ is a commutative selector function that is total and associative on $\Sigma^n - B^{=n}$ it follows from Proposition 3.2 that $E'_n$ is a transitive relation. Hence it follows from Proposition 3.5 that $G'_n$ contains a nonempty strong distance-one target-clique, i.e., an induced subgraph $T = (V_T, E_T)$, $V_T \neq \emptyset$, having the following properties:

1. $T$ is a maximal strong clique.

2. For all $u \in V_T$ and for all $v \in (\Sigma^n - B^{=n}) - V_T$, $(u,v) \notin E'_n$ and $(v,u) \in E'_n$.

Let $v_n$ denote the lexicographically largest string in $V_T$. (This is just for specificity. Any string from $V_T$ would function the same way.) It follows that for $n \in \mathbb{N}$, if $B^{=n} \neq \emptyset$ and $\Sigma^n - B^{=n} \neq \emptyset$, then there exists a string $v_n \in \Sigma^n - B^{=n}$ such that for all $x \in \Sigma^n$,

$$x \in B \iff v_n \notin \text{set-}f(x, v_n).$$

Define, for all $n \in \mathbb{N}$,

$$h'(1^n) = \begin{cases} 0^{n+1} & \text{if } B^{=n} = \emptyset \\ 01^n & \text{if } B^{=n} = \Sigma^n \\ 1v_n & \text{if } B^{=n} \neq \emptyset \text{ and } B^{=n} \neq \Sigma^n. \end{cases}$$

Note that $h'$ on any input $1^n$ outputs a string of length exactly $n+1$. Let $L' = \{\langle x, 1y \rangle \mid |x| = |y| \wedge y \notin \text{set-}f(x,y)\} \cup \{\langle x, 01^{|x|} \rangle \mid x \in \Sigma^*\}$. Clearly, $L' \in \text{coNP}$. Due to the above construction we also have, for all $x \in \Sigma^*$,

$$x \in B \iff \langle x, h'(1^{|x|}) \rangle \in L'.$$

This shows that $B \in \text{coNP}/n+1$. ❏

Since as noted earlier $(\text{P}/1) - \text{NPMV-sel} \neq \emptyset$, Theorem 5.9 yields the following corollary.

**Corollary 5.10**  1. A-NP-sel $\subsetneq (\text{NP} \cap \text{coNP})/n+1$.

2. $\text{A}_\ell$-NPSV-sel $\subsetneq \text{NP}/n+1 \cap \text{coNP}/n+1$.

3. A-NPMV$_\text{t}$-sel $\subsetneq \text{NP}/n+1 \cap \text{coNP}/n+1$.

4. $\text{A}_\ell$-NPMV$_\text{t}$-sel $\subsetneq \text{NP}/n+1 \cap \text{coNP}/n+1$.

We mention in passing that—since $\log_2((2^{n+1} - 1) + 1) = n+1$ and thus one can include in the advice-seeking s-tournament all lengths up to the current one—the $n+1$ advice bounds in part 1 of Corollaries 5.4 and 5.10 even hold in the "strong advice" model of Ko, Balcázar, and Schöning [Ko87, BS92] in which advice must work for all strings up to and including a given length (see also the discussion in [HT96]). The $n+1$ advice bound in the "NP$/n+1$" of part 3 of Corollary 5.10 for the same reason holds also in the "strong advice" model. Since the "coNP$/n+1$" part of the proof of part 2 of Theorem 5.9 had *two* special-meaning advice cases, at first it might seem that regarding the "coNP$/n+1$" of part 3 of Corollary 5.10 $\lceil \log_2((2^{n+1} - 1) + 2) \rceil = n+2$ advice bits, rather than $n+1$ advice bits, are needed in the "strong advice" model. However, note that we can hardcode into our machine the issue of whether some one string, say $\epsilon$, is in our A-NPMV$_\text{t}$-sel set, and then can leave $\epsilon$ out of all the tournaments and always handle it directly. This ensures that the size of the collection of strings that we have to worry about of length up to $n$ is $2^{n+1} - 2$, and so the advice size we need is $\lceil \log_2((2^{n+1} - 2) + 2) \rceil = n+1$. So A-NPMV$_\text{t}$-sel indeed is in NP$/_\text{strong}\, n+1 \cap$ coNP$/_\text{strong}\, n+1$, where by $/_\text{strong}$ we mean the strong advice model. Keeping in mind that $(\text{P}/1) - \text{NPMV-sel} \neq \emptyset$, we state the discussion of this paragraph as follows.

**Proposition 5.11**  1. A-P-sel $\subsetneq$ P $/_\text{strong}\, n+1$.

2. A-NP-sel $\subsetneq (\text{NP} \cap \text{coNP}) /_\text{strong}\, n+1$.

3. A-NPMV$_\text{t}$-sel $\subsetneq$ NP $/_\text{strong}\, n+1 \cap$ coNP $/_\text{strong}\, n+1$.

It is natural to wonder whether there is some natural "NPSV" analog of part 1 of Theorem 5.9 that yields an $(\text{NP} \cap \text{coNP})/n + 1$ result. In light of part 2 of Theorem 5.8, one in particular might hope that $\text{A}_\ell\text{-NPSV-sel} \cap \text{NP} \subseteq (\text{NP} \cap \text{coNP})/n + 1$. However, we see no way of proving this. The key obstacle is obtaining the "$\text{NP} \cap \text{coNP}$" of "$(\text{NP} \cap \text{coNP})/n + 1$" when dealing with partial function classes such as NPSV. One can patch one's way around the partialness of NPSV, namely, by using the advice to give not just an appropriate (via part 3 of Proposition 3.3) string from the set but also a certificate proving that the string is in the set. However, doing so requires enough bits to encode the certificate. So what one obtains, if one modifies part 1 of Theorem 5.9 carefully via the strategy just described, is the following result, in which $\text{NP}_1$ denotes the class, introduced by Kintala and Fischer in 1977 ([KF77], see also for contrast [KF80]), of sets that are accepted by NP machines that use at most a linear number of nondeterministic moves.

**Theorem 5.12** $\text{A}_\ell\text{-NPSV-sel} \cap \text{NP}_1 \subseteq (\text{NP} \cap \text{coNP})/\mathcal{O}(n)$.

Crucial in seeing that Theorem 5.12 holds is the following. When given advice asserting that our $\text{A}_\ell$-NPSV-selective set, $B$, contains $y$ as a "magic" (in the sense of part 3 of Proposition 3.3) string for length $|y|$, only if the advice contains a correct certificate for $y \in B$ do we go forward and run our NPSV-selector function $f$, i.e., $f(x, y)$, where $x$, $|x| = |y|$, is the string whose membership we are interested in. So we run $f$ only on cases on which at least one of the inputs on which we run $f$ is clearly known by us to belong to $B$. By the definition of NPSV-selectivity, $f$ on such inputs will always be defined. Thus, working hand in hand with the advice and the definition of NPSV-selectivity, we have in effect locally total-ized $f$, and this is what creates an $\text{NP} \cap \text{coNP}$ set, and yields the $(\text{NP} \cap \text{coNP})/\mathcal{O}(n)$ result.

To the best of our knowledge, Theorem 5.12 is only the second time in the literature that the direct local totalization of NPSV functions appears. The first time was in Hemaspaandra et al. [HNOS96b], though in contrast there they were in the context of a Ko-type [Ko83] advice setting, i.e., one in which (ignoring certificate space) a linear number of linear-sized strings were being provided. Here, we have given a second application of their local totalization technique, this time in the context of the (ignoring certificate space) one-string advice of the sort provided by part 3 of Proposition 3.3.

Finally, we would like to remark on results involving weak associativity. If we assume the NPSV- or NPMV-selectors to be only weakly associative, we are merely able to show containment in $\text{NP}/n+1$.

## 6 Printability and Nonimmunity

Associativity yields additional simplicity properties. Let us consider nonimmunity results (i.e., presence of infinite subsets). For general NPSV-selective sets and thus also for NP- and P-selective sets, Theorem 6.1 holds. In contrast, for associatively NPSV-selective sets, we have Theorem 6.2. To see how these bounds relate, note that $\text{UP}^{\text{NP}} \subseteq \text{NP}^{\text{NP}}$. Of course, $\text{NP}^{\text{NP}}$ is synonymous with $\Sigma_2^p$; however, so as to ensure the relationship to $\text{UP}^{\text{NP}}$—and the way we'll use those "NP"s in our proofs—shows clearly, throughout this section we will write $\text{NP}^{\text{NP}}$ rather than $\Sigma_2^p$.

**Theorem 6.1** *Every infinite NPSV-selective set $B$ has an infinite $\text{FP}^{B \oplus \text{NP}^{\text{NP}}}$-printable subset.*

**Theorem 6.2** *Every infinite set $B$ that is either A-NPSV-selective or $\text{A}_\ell$-NP-selective has an infinite $\text{FP}^{B \oplus \text{UP}^{\text{NP}}}$-printable subset.*

**Corollary 6.3**   *1. Every infinite P-selective set $B$ has an infinite $\text{FP}^{B \oplus \text{NP}^{\text{NP}}}$-printable subset.*

2. Every infinite NP-selective set $B$ has an infinite $\text{FP}^{B \oplus \text{NP}^{\text{NP}}}$-printable subset.

3. Every infinite A-P-selective (or even $A_\ell$-P-selective) set $B$ has an infinite $\text{FP}^{B \oplus \text{UP}^{\text{NP}}}$-printable subset.

4. Every infinite A-NP-selective set $B$ has an infinite $\text{FP}^{B \oplus \text{UP}^{\text{NP}}}$-printable subset.

**Proof of Theorem 6.1:** Let $B$ be an infinite NPSV-selective set. By Fact 2.3 there exists a commutative NPSV-selector $f$ for $B$. Let $n \in \mathbb{N}$. Consider the $f$-induced digraph on $B^{=n}$ (in the sense defined in Section 3), call it $G_{B,n,f}$. Since $f$ is always defined when at least one of its inputs is in $B$, it follows that $G_{B,n,f}$ is an s-tournament. By a standard result (stated as Theorem 3.1 in [HO02]; the theorem refers to tournaments rather than s-tournaments but the difference does not affect this result) of Ko [Ko83] about tournament theory, applied to $G_{B,n,f}$, we have that for each $n \in \mathbb{N}$ there exists a set $D_n \subseteq B^{=n}$ such that

1. $\|D_n\| \leq n + 1$ and

2. for every string $x \in B^{=n}$, there exists a string $y \in D_n$ such that $y \in \text{set-}f(x, y)$.

Of course, when $B^{=n} = \emptyset$, $G_{B,n,f}$ has no nodes and $D_n = \emptyset$.

Since $f$ is a selector for $B$ and thus $f(u,v) = v$ for all $u \notin B$ and all $v \in B$ we can state that for each $n \in \mathbb{N}$ such that $B^{=n} \neq \emptyset$ there exists a set $D_n \subseteq B^{=n}$ such that

1. $\|D_n\| \leq n + 1$ and

2. for every string $x \in \Sigma^n$, there exists a string $y \in D_n$ such that $y \in \text{set-}f(x, y)$.

Let *setcode* denote some standard encoding (of finite sets into strings) that is computable and invertible in polynomial time. Let $\langle \cdot, \cdot \rangle$ be a standard polynomial-time computable, polynomial-time invertible pairing function. Define

$$E = \{\langle 1^n, \textit{setcode}(\{y_1, y_2, \ldots, y_j\})\rangle \mid 1 \leq j \leq n + 1 \wedge$$
$$|y_1| = \cdots = |y_j| = n \wedge (\forall x \in \Sigma^n)(\exists i : 1 \leq i \leq j)[x \notin \textit{set-}f(x, y_i)]\}.$$

Observe that the condition $x \notin \textit{set-}f(x, y_i)$ is equivalent to $\textit{set-}f(x, y_i) \in \{\emptyset, \{y_i\}\}$. Clearly, $E \in \text{coNP}$. For all $n$, let $g(1^n)$ denote the lexicographically largest string $\omega$ such that $\langle 1^n, \omega \rangle \in E$ if such a string $\omega$ exists. (Note that in light of the discussion earlier in this proof and the properties and single-valuedness of $f$, it holds that, for each $n$ such that $B^{=n} \neq \emptyset$, there exists at least one string $w$ such that $\langle 1^n, w \rangle \in E$.) Otherwise $g(1^n)$ is undefined. Given $1^n$ one can by applying a prefix search procedure compute $g(1^n)$ in polynomial time with the help of oracle queries to the oracle $F = \{\langle 1^n, z \rangle \mid n \in \mathbb{N} \wedge z \in \Sigma^* \wedge (\exists \omega \in \Sigma^*)[\langle 1^n, z\omega \rangle \in E]\}$. Clearly, $F \in \text{NP}^{\text{NP}}$ and hence $g \in \text{FP}^{\text{NP}^{\text{NP}}}$.

We claim that, for all $n$, if $B^{=n} \neq \emptyset$ then for every $\langle 1^n, \textit{setcode}(\{y_1, y_2, \ldots, y_j\})\rangle \in E$ ($1 \leq j \leq n + 1$)—and so in particular for $\langle 1^n, g(1^n)\rangle$—there exists an index $i$, $1 \leq i \leq j$, such that $y_i \in B^{=n}$. To see this, suppose that $B^{=n} \neq \emptyset$ yet there exists some $\langle 1^n, \textit{setcode}(\{y_1, y_2, \ldots, y_j\})\rangle \in E$ ($1 \leq j \leq n + 1$) such that, for all $i$, $1 \leq i \leq j$, $y_i \notin B^{=n}$. Let $x \in B^{=n}$. Due to the definition of $E$ we have that there exists an index $i$, $1 \leq i \leq j$, such that either $\textit{set-}f(x, y_i) = \{y_i\}$ or $f(x, y_i)$ is undefined, both of which are impossible since $f$ is an NPSV-selector for $B$ and so, when exactly one of its arguments ($x$ in this case) is in $B$, it outputs that argument.

Consider the set

$$C = B \cap \bigcup_{\{i \in \mathbb{N} \mid g(1^i) \text{ is defined}\}} \textit{setcode}^{-1}(g(1^i)).$$

Note that $C$ is infinite if $B$ is infinite. Also $C \subseteq B$. In order to see that $C$ is $\text{FP}^{B \oplus \text{NP}^{\text{NP}}}$-printable, consider the following DPTM $M$: On input $1^n$, $M(1^n)$ computes $g(1^0), g(1^1), \ldots, g(1^n)$ with the help of the $\text{NP}^{\text{NP}}$ part, $F$, of its $B \oplus \text{NP}^{\text{NP}}$ oracle. It then computes the polynomial-sized set $\bigcup_{\{i \leq n \mid g(1^i) \text{is defined}\}} setcode^{-1}(g(1^i)))$ and then, using the $B$ part of its oracle, finds which of these strings are in $B$ and outputs them. ❑

Theorem 6.1 should be contrasted with the result of Hemaspaandra, Ogihara, Zaki, and Zimand [HOZZ04] that P-sel is not (weak-$\text{FP}^{\text{NP}^{\text{NP}}}$-rankable)-immune.

**Proof of Theorem 6.2:** In light of part 3 of Corollary 4.7, it actually holds that $\text{A}_\ell$-NP-sel $\supseteq$ A-NP-sel = A-NPSV-sel. Hence it suffices to show the claim for $\text{A}_\ell$-NP-selective sets. Let $B$ be an infinite $\text{A}_\ell$-NP-selective set. By part 2 of Theorem 4.2, $\text{A}_\ell$-NP-sel = $\text{A}_\ell$C-NP-sel. So, let $f$ be a commutative $\text{NPSV}_t$-selector for $B$ that is associative at each length.

Consider the function $score$ that is defined as $score(x) = \|\{z \in \Sigma^{|x|} \mid f(x, z) = x\}\|$. Note that for each $n \in \mathbb{N}$ and for each $x \in \Sigma^n$, (a) $x \notin B \implies score(x) \leq 2^n - \|B^{=n}\|$ and (b) $x \in B \implies score(x) \geq 2^n - \|B^{=n}\| + 1$. So for all $x \in B$ and all $y \notin B$ such that $|x| = |y|$, $score(y) < score(x)$. It follows from the fact that $f$ is commutative, total, single-valued, and length-associative that for all $n$, $\max\{score(y) \mid |y| = n\} = 2^n$ (consider Proposition 3.3 applied to the $f$-induced digraph on the nodes $(\Sigma^*)^{=n}$). So for each $n$ there exists exactly one string at length $n$, call it $d_n$, such that $score(d_n) = \max\{score(y) \mid |y| = n\} = 2^n$. Thus, the set

$$F = \{\langle 1^n, z \rangle \mid n \in \mathbb{N} \land z \in (\Sigma^*)^{\leq n} \land (\exists \omega : |\omega| = n - |z|)[score(z\omega) = 2^n]\}$$

is in $\text{UP}^{\text{NP}}$. (To see this, note that $score(x) = 2^{|x|}$ is in our setting equivalent to $(\forall y \in \Sigma^{|x|})[x = y \lor f(x, y) \neq y]$.) Given $1^n$, computing $d_n$ can be done in polynomial time with the help of oracle $F$. Let $C = B \cap \{d_i \mid i \geq 0\}$. Clearly, since $B$ is infinite, by our choice of $d_i$ we have that $C$ is both infinite and $\text{FP}^{B \oplus \text{UP}^{\text{NP}}}$-printable. ❑

So, Theorems 6.1 and 6.2 have been proven.[6]

It follows from the proof of Theorem 6.2 that any associatively NPSV selective set for which there is an infinite set of lengths at which we know it is not empty is in fact not coNP-immune. This is a *very* weak type of partial census information.

**Definition 6.4** *We say a set $B$ is hintable if there is an infinite tally set, $T \subseteq 1^*$, such that $T \in \text{P}$ and*

$$(\forall i)[1^i \in T \implies \|B^{=i}\| \neq 0].$$

**Theorem 6.5** *Any hintable A-NPSV-selective (or even $\text{A}_\ell$-NPSV-selective) set has an infinite coNP subset (i.e., languages in $\text{A}_\ell$-NPSV-sel $\cap$ Hintable are not coNP-immune).*

**Proof:** Let $B$ be an infinite hintable $\text{A}_\ell$-NPSV-selective set. Let the tally set $T$ be a hint set for $B$ in the sense of Definition 6.4. Let $f$ be a commutative NPSV-selector for $B$ that is total at each length $n$ such that $B^{=n} \neq \emptyset$ and associative at each length $n$ such that $B^{=n} \neq \emptyset$. Such a selector for $B$ exists by Corollary 4.7. Define the function $score$ as in the proof of Theorem 6.2. Since $B$ is $\text{A}_\ell$-NPSV-selective, we have that, for all $n$, if $B^{=n} \neq \emptyset$ then there exists a unique string of length $n$ having a score of $2^n$ and it is in $B$. (If $B^{=n} = \emptyset$ then there might be no string at length $n$ having a score of $2^n$.) The set
$$C = \{x \mid 1^{|x|} \in T \land score(x) = 2^n\}$$
is an infinite subset of $B$ and is in coNP since it will hold that $C = \{x \mid 1^{|x|} \in T \land (\forall y : |y| = |x|)[y = x \lor y \notin set\text{-}f(x,y)]\}$. ❑

---

[6]If one cares about the number of queries to $B$ needed in the printability claims of Theorems 6.1 and 6.2 and Corollary 6.3, the following, which was communicated to us (regarding a somewhat weaker set of results in an earlier version of this paper) by Till Tantau [Tan01], can be observed: One can limit one's queries to $B$ to a logarithmic number via putting the queries to $B$ into a Toda-like ordering ([Tod91], see also [HT03]) and binary searching to find which are in and which are out. (For the nondeterministic selectivity claims, one will use the power of NP in the oracle to Toda-order the strings. Also, the partial-function cases do not present a problem, e.g., on those one can modify for the partial case the "bubblesort"-type proof given by Hemaspaandra and Torenvliet [HT03, Proof of Lemma 4.5].

## 7 Are All P-Selective Sets Associatively P-Selective?

Many of our results give simplicity properties of A-P-sel. It is natural to wonder whether in fact A-P-sel is all of P-sel, that is, whether all P-selective sets are associatively P-selective. In this section, we prove upper bounds on the power needed to associatively select sets. One consequence, Corollary 7.3, is that if P = NP, then A-P-sel and P-sel are equal.

**Theorem 7.1**  1. *Every* $\text{NPMV}_t$-*selective set has a single-valued, commutative selector function in* $\text{FP}_t^{\text{NP}}$ *that is associative.*

2. *Every* NPMV-*selective set $A$ has a single-valued, commutative selector function in* $\text{FP}_t^{\text{NP}}$ *that is associative on $A$.*

3. *Every* NPMV-*selective set has a single-valued, commutative selector function in* $\text{FP}_t^{\Sigma_2^p}$ *that is associative.*

**Proof:** We first prove part 1 of the result. Let $A$ be an $\text{NPMV}_t$-selective set. Let $f$ be an $\text{NPMV}_t$-selector for $A$. Without loss of generality, assume $f$ to be commutative (see Fact 2.3). For every pair of strings $x$ and $y$, $\omega$ is called a connector of $x$ and $y$ if and only if either (a) $\omega \in set\text{-}f(x,\omega)$ and $y \in set\text{-}f(\omega, y)$ or (b) $x \in set\text{-}f(x, \omega)$ and $\omega \in set\text{-}f(\omega, y)$. Note that for any two strings $x$ and $y$, there always exist connectors of $x$ and $y$, namely $x$ and $y$. Define the function $g$ by setting, for all $x$ and $y$,

$$g(x,y) = \begin{cases} y & \text{if the lexicographically smallest connector } \omega \text{ of } x \text{ and } y \text{ yields } \\ & \omega \in set\text{-}f(x,\omega) \land y \in set\text{-}f(\omega,y) \text{ but not } x \in set\text{-}f(x,\omega) \land \omega \in set\text{-}f(\omega,y) \\ x & \text{if the lexicographically smallest connector } \omega \text{ of } x \text{ and } y \text{ yields } \\ & x \in set\text{-}f(x,\omega) \land \omega \in set\text{-}f(\omega,y) \text{ but not } \omega \in set\text{-}f(x,\omega) \land y \in set\text{-}f(\omega,y) \\ \max(x,y) & \text{if the lexicographically smallest connector } \omega \text{ of } x \text{ and } y \text{ yields } \\ & \omega \in set\text{-}f(x,\omega) \land y \in set\text{-}f(\omega,y) \text{ and } x \in set\text{-}f(x,\omega) \land \omega \in set\text{-}f(\omega,y). \end{cases}$$

Note that $g$ is a single-valued function in $\text{FP}_t^{\text{NP}}$. To compute $g(x, y)$ the NP oracle is first used to—via either binary search or prefix search—compute the lexicographically smallest connector of $x$ and $y$, and then via two more queries is used to determine which of the three cases from the definition of $g$ is applicable. Note that $g$ is commutative and self-contained.

To see that $g$ is even a selector for $A$, suppose that there are strings $x$ and $y$ such that $x \in A$ and $y \notin A$ yet $g(x, y) = y$. Let $\omega$ be the lexicographically smallest connector for $x$ and $y$. It follows that either (a) $\omega \in set\text{-}f(x,\omega) \land y \in set\text{-}f(\omega,y)$ but not $x \in set\text{-}f(x,\omega) \land \omega \in set\text{-}f(\omega,y)$ or (b) $y = \max(x,y)$, $\omega \in set\text{-}f(x,\omega) \land y \in set\text{-}f(\omega,y)$, and $x \in set\text{-}f(x,\omega) \land \omega \in set\text{-}f(\omega,y)$. However, in both cases the properties of a selector function (in this case $f$) imply that if $x \in A$ so are $\omega$ and $y$, contradicting our assumption.

It remains to show that $g$ is associative. Suppose that $g$ is not associative. Let $a, b, c$ be a counterexample to the associativity of $g$. Since $g$ is a single-valued, commutative self-contained function, it follows from Corollary 3.4 that both $||\{a, b, c\}|| = 3$ and the $g$-induced digraph on $a, b, c$ is a 3-cycle (plus three self-loops). Without loss of generality, suppose that $g(a, b) = b$, $g(b, c) = c$, and $g(a, c) = a$. Let $\omega$ be the lexicographically smallest connector among all connectors for $a$ and $b$, for $b$ and $c$, and for $a$ and $c$.

Suppose that $\omega$ is a connector for $a$ and $b$. Due to symmetry we can make that assumption without loss of generality. Since $g(a, b) = b$ there are two possible scenarios.

**Case 1** $\omega \in \text{set-}f(a, \omega) \wedge b \in \text{set-}f(\omega, b)$ but not $a \in \text{set-}f(a, \omega) \wedge \omega \in \text{set-}f(\omega, b)$.

Since $f$ is total, $\text{set-}f(\omega, c) \neq \emptyset$. If $\text{set-}f(\omega, c) = \{c\}$ then $\omega$ is also a connector for $a$ and $c$ contradicting, in light of the definition of $g$ and the fact that in this case $\omega \in \text{set-}f(a, \omega)$ and $\text{set-}f(\omega, c) = \{c\}$ yet $\omega \notin \text{set-}f(\omega, c)$, $g(a, c) = a$. If $\text{set-}f(\omega, c) = \{\omega\}$ then $\omega$ is also a connector for $b$ and $c$ in such a way as to similarly contradict $g(b, c) = c$. If $\text{set-}f(\omega, c) = \{\omega, c\}$, there are two possibilities. The first is that $\text{set-}f(\omega, b) = \{b\}$. However, that implies $g(b, c) = b$, which contradicts our assumption that $g(b, c) = c$. The second possibility is that $\text{set-}f(\omega, b) = \{b, \omega\}$. This implies (under the rules of Case 1 and the subcase we are in) that $\text{set-}f(a, \omega) = \{\omega\}$. However, this implies that $g(c, a) = c$, which contradicts our assumption that $g(a, c) = a$.

**Case 2** $b = \max(a, b)$, $\omega \in \text{set-}f(a, \omega) \wedge b \in \text{set-}f(\omega, b)$, and $a \in \text{set-}f(a, \omega) \wedge \omega \in \text{set-}f(\omega, b)$.

Since $f$ is total, $f(\omega, c)$ is defined. If $\text{set-}f(\omega, c) = \{c\}$ then we have $g(a, c) = c$, contradicting our assumption that $g(a, c) = a$. If $\text{set-}f(\omega, c) = \{\omega\}$ then we have $g(b, c) = b$, contradicting our assumption that $g(b, c) = c$. If $\text{set-}f(\omega, c) = \{\omega, c\}$ then the definition of $g$ and the assumed values of $g$ imply that $b = \max(a, b)$, $a = \max(a, c)$, and $c = \max(b, c)$, which contradicts $||\{a, b, c\}|| = 3$.

It follows that $g$ is associative which completes the proof of part 1.

We next show part 2. This can be shown quite similarly and so we will only sketch the main differences to the proof of part 1. Let $A$ be an NPMV-selective set. Let $f$ be an NPMV-selector for $A$. Without loss of generality, assume $f$ to be commutative (Fact 2.3). Observe that $f(x, y)$ and also one of $f(x, x)$ and $f(y, y)$ is defined whenever at least one of $x$ and $y$ is in $A$. Thus, whenever at least one of $x$ and $y$ is in $A$ there exists a connector $z$ of $x$ and $y$ such that $z \leq_{lex} \max(x, y)$. Define the function $h$ by setting, for all $x$ and $y$,

$$h(x, y) = \begin{cases} y & \text{if } x \text{ and } y \text{ have a connector } z \text{ such that } z \leq_{lex} \max(x, y) \text{ and the lexicographically smallest connector } \omega \text{ of } x \text{ and } y \text{ yields } \omega \in \text{set-}f(x, \omega) \wedge y \in \text{set-}f(\omega, y) \text{ but not } x \in \text{set-}f(x, \omega) \wedge \omega \in \text{set-}f(\omega, y), \\ x & \text{if } x \text{ and } y \text{ have a connector } z \text{ such that } z \leq_{lex} \max(x, y) \text{ and the lexicographically smallest connector } \omega \text{ of } x \text{ and } y \text{ yields } x \in \text{set-}f(x, \omega) \wedge \omega \in \text{set-}f(\omega, y) \text{ but not } \omega \in \text{set-}f(x, \omega) \wedge y \in \text{set-}f(\omega, y), \\ \max(x, y) & \text{if } x \text{ and } y \text{ have a connector } z \text{ such that } z \leq_{lex} \max(x, y) \text{ and the lexicographically smallest connector } \omega \text{ of } x \text{ and } y \text{ yields both } \omega \in \text{set-}f(x, \omega) \wedge y \in \text{set-}f(\omega, y) \text{ and } x \in \text{set-}f(x, \omega) \wedge \omega \in \text{set-}f(\omega, y), \\ \max(x, y) & \text{if } x \text{ and } y \text{ have no connector } z \text{ such that } z \leq_{lex} \max(x, y). \end{cases}$$

Note that $h$ is a single-valued $\mathrm{FP}_t^{\mathrm{NP}}$ function that is commutative. Similarly to the proof of part 1, one can now show that $h$ is a selector function for $A$ that is associative on $A$. For the latter note that any connector $z$ of two strings $x, y \in A$ is also in $A$.

Part 3 follows from the proofs of parts 1 and 2. The function $h$ as defined in the proof of part 2 is a total, commutative, and single-valued $\mathrm{FP}^{\mathrm{NP}}$ function. We can now repeat the entire proof of part 1 but replacing $f$ by $h$ and $\mathrm{NPMV}_t$ by $\mathrm{FP}_t^{\mathrm{NP}}$ in that proof. It follows that the function $g$ as defined in the proof of part 1 is now in our modified version a total single-valued $\mathrm{FP}^{\Sigma_2^p}$-selector for $A$ that is commutative and associative. ❏

**Corollary 7.2**  1. Every P-selective set, and even every FP-selective set, has a commutative selector function in $\mathrm{FP}_t^{\mathrm{NP}}$ that is associative.

2. Every NP-selective set has a commutative selector function in $\mathrm{FP}_t^{\mathrm{NP}}$ that is associative.

3. Every NPSV-selective set $A$ has a commutative selector function in $\mathrm{FP}_t^{\mathrm{NP}}$ that is associative on $A$.

4. Every NPSV-selective set has a commutative selector function in $\mathrm{FP}_t^{\Sigma_2^p}$ that is associative.

The above corollary is a straightforward consequence of Theorem 7.1. Regarding the "every FP-selective" in part 1, recall FP-sel = P-sel from Fact 2.3.

**Corollary 7.3** *If* P = NP *then* P-sel = A-P-sel, FP-sel = A-FP-sel, NP-sel = A-NP-sel, NPSV-sel = A-NPSV-sel, $\mathrm{NPMV}_t$-sel = A-$\mathrm{NPMV}_t$-sel, *and* NPMV-sel = A-NPMV-sel.

The above corollary follows directly from Theorem 7.1 and and Corollary 7.2. Regarding the NPSV-sel = A-NPSV-sel and NPMV-sel = A-NPMV-sel conclusions of the above corollary, recall that P = NP $\iff$ P = $\Sigma_2^p$.

Theorem 5.3 provides a sufficient condition, based on an algebraic property for selector functions, for P-sel $\subseteq$ P/$\mathcal{O}(n)$. If one were interested only in *structural complexity-class-collapse* sufficient conditions, Theorem 5.3 would be no improvement over the P = NP sufficient condition implicit in the result of Hemaspaandra and Torenvliet that P-sel $\subseteq$ NP/$\mathcal{O}(n)$ (and thus P-sel $\subseteq$ P/$\mathcal{O}(n)$ if P = NP), since in light of Corollary 7.3 the best known structural complexity-class-collapse condition sufficient to imply P-sel = A-P-sel is also the collapse P = NP. However, we feel that this is the wrong view, and that "P-sel = A-P-sel" is probably a fundamentally different type of assumption than P = NP. For example, if P-sel = A-P-sel were in fact equivalent to P = NP, then P = NP would (trivially) be not just a sufficient condition for P-sel = A-P-sel but also would be a necessary condition. In fact, not only is P = NP not known to be necessary for P-sel = A-P-sel, but in fact no structural complexity-class-collapse condition—not even very weak collapses like P = UP or P = ZPP—is known to be necessary. Our point here is that, though by Corollary 7.3 P = NP is one way to achieve P-sel = A-P-sel, we conjecture that it is unlikely that one can prove that it characterizes P-sel = A-P-sel. And thus, our P-sel = A-P-sel sufficient condition for P-sel $\subseteq$ P/$\mathcal{O}(n)$ is best viewed as a new algebraic sufficient condition quite different from the known (and extremely demanding) structural complexity-class-collapse sufficient conditions.

We do have a structural complexity-class-collapse condition, namely the collapse of the polynomial hierarchy, that follows from the assumption that every NPMV-selective set is in fact associatively NPMV-selective. ($\mathrm{S}_2$ is the symmetric alternation class of Canetti [Can96] and Russell and Sundaram [RS98]. Note that a collapse to $\mathrm{S}_2^{\mathrm{NP}}$ is known to be at least as strong as a collapse to $\mathrm{ZPP}^{\Sigma_2^p}$ or to $\Sigma_3^p$, since $\mathrm{S}_2^{\mathrm{NP}} \subseteq \mathrm{ZPP}^{\Sigma_2^p} \subseteq \Sigma_3^p$ ([CCHO03], see that paper also for the definition of $\mathrm{S}_2^{\mathrm{NP}}$).)

**Theorem 7.4** NPMV-sel = A-NPMV-sel $\implies$ PH = $S_2^{NP}$.

**Proof:** Assume NPMV-sel = A-NPMV-sel. By Corollary 5.10 we have that A-NPMV-sel $\subseteq$ NP/$n+1$ $\cap$ coNP/$n+1$. So under our assumption NPMV-sel $\subseteq$ NP/$n+1$ $\cap$ coNP/$n+1$. However, it is well-known that NP $\subseteq$ NPMV-sel. Putting all the pieces together we obtain NP $\subseteq$ coNP/$n+1$. This itself, by Cai et al.'s [CCHO03] recent strengthening of Yap's Theorem [Yap83], implies that PH = $S_2^{NP}$ (equivalently, PH = $NP^{NP^{NP}}$ = $ZPP^{NP^{NP}}$ = $S_2^{NP}$). ❑

Finally, we mention two open issues. First, can one prove a complete or partial converse of other parts of Corollary 7.3? The second issue is the following. One would ultimately like to know whether all P-selective sets have linear deterministic advice, i.e., whether P-sel $\subseteq$ P/$\mathcal{O}(n)$. This paper gives a new sufficient condition for that, namely, P-sel $\subseteq$ P/$\mathcal{O}(n)$ if all P-selective sets have associative (or even merely length-associative) selector functions.

**Acknowledgments:** We are deeply grateful to Klaus Wagner for his valuable suggestions. We thank the COCOON 2001 and SICOMP referees for their helpful comments, and we thank Dieter Kratsch, Haiko Müller, Till Tantau, and Leen Torenvliet for helpful conversations and comments.

[Tod91]   S. Toda. On polynomial-time truth-table reducibilities of intractable sets to P-selective sets. *Mathematical Systems Theory*, 24(2):69–82, 1991.

[Yap83]   C. Yap. Some consequences of non-uniform conditions on uniform classes. *Theoretical Computer Science*, 26(3):287–300, 1983.

## A   Exponential-Time Selector Functions

This appendix notes that all P-selective sets have commutative, associative selector functions computable in exponential—by which we mean *linearly* exponential—time.

We first prove a result that is far weaker than results proven in Section 7, but that we'll refer to at one point later in this section.

**Fact A.1** P = PP $\implies$ P-sel = $A_\ell$-P-sel.

**Proof:** Suppose P = PP. Let $B \in$ P-sel via P-selector function $f'$. By Fact 2.3, there is a commutative P-selector function $f$ for $B$. We would like to show that there exists a P-selector $g$ for $B$ that is associative at each length.

Let the definition of the function *score* be as in the proof of Theorem 6.2. Recall that for all $x \in B$ and all $y \notin B$ such that $|x| = |y|$, $score(y) < score(x)$. We define $g$ as follows, for all $x, y \in \Sigma^*$,

$$g(x,y) = \begin{cases} f(x,y) & \text{if } |x| \neq |y| \\ x & \text{if } |x| = |y| \text{ and } score(x) > score(y) \\ y & \text{if } |x| = |y| \text{ and } score(x) < score(y) \\ \max(x,y) & \text{otherwise.} \end{cases}$$

$g$ is associative at each length. It remains to show that $g$ is indeed a P-selector for $B$. For this it suffices to show that for equal-length strings $x$ and $y$, $g$ has the properties of a P-selector function for $B$.

So, let $x$ and $y$ be of the same length, say $|x| = |y| = n$. Note that when either $x, y \in \left(\overline{B}\right)^{=n}$ or $x, y \in B^{=n}$, a P-selector for $B$ may safely output either $x$ or $y$. Also, $g(z, z)$ correctly equals $z$ for all strings $z$. Finally, if $x \neq y$ and $\|\{x, y\} \cap B^{=n}\| = 1$ then by the fact about *score* noted above, $score(x) \neq score(y)$, and the string that is in the set has the larger function value with respect to the function *score*. Thus, the constructed function $g$ is indeed a selector function for $B$. Since P = PP we can make the case distinction in polynomial time, and so $g$ is clearly computable in polynomial time if P = PP, and so $g$ is a P-selector function for $B$ if P = PP. ❑

The proof of Fact A.1 in fact actually establishes that every P-selective set has an $FP^{PP}$-selector function that is associative at each length.

What does it take to achieve associative selectivity? The following result shows that every P-selective set has a commutative, associative selector function computable in exponential time. We say a set $B$ belongs to ETIME-CA-sel if and only if there exists a commutative, associative selector function for $B$ that is computable in time $2^{\mathcal{O}(n)}$.

**Theorem A.2** P-sel $\subseteq$ ETIME-CA-sel.

**Proof:** Let $B$ be any P-selective set and let $g$ be a P-selector for $B$. Without loss of generality assume that $g$ is commutative. As a convention, for any total order *ord* let $\max_{ord}(x, y)$ denote the maximum of $x$ and $y$ with respect to the order *ord* ($\preceq_{ord}$). Of course, $\max_{ord}(x, x) = x$. We will, inductively, define total orders $S_0, S_1, S_2, \ldots$ such that

1. for all $i$, $S_i$ is a total order of the strings in $\Sigma^{*\leq i}$,

2. each $S_i$, $i \geq 1$, respects $S_{i-1}$,

3. the function
$$f(x,y) = \max_{S_{\max(|x|,|y|)}}(x,y) = \begin{cases} x & \text{if } x \succeq y \text{ in } S_{\max(|x|,|y|)} \\ y & \text{otherwise,} \end{cases}$$
is an associative selector function for $B$, and

4. $f$ is computable in deterministic time $2^{\mathcal{O}(n)}$.

We will represent any total order *ord* over a finite set of strings $E$, $\|E\| = k$, in the form of a list $(x_1, x_2, x_3, \ldots, x_k)$ where $x_i \preceq_{ord} x_j$ if and only if $i \leq j$.

By the comment after Fact A.1 and, since the approach of the proof of Theorem 4.2 also works for $\text{FP}^{\text{PP}}$-selector functions, we know that $B$ has a commutative $\text{FP}^{\text{PP}}$-selector $h$ that is associative at each length. In fact, looking back at the proof of the earlier theorem, that selector function works completely based on taking the ranks among $2^n$ strings, and thus, though some functions in $\text{FP}^{\text{PP}}$ may potentially require time $2^{n^{\mathcal{O}(1)}}$, this particular function clearly can be computed in deterministic time $2^{\mathcal{O}(n)}$. This is important in light of our ETIME overall goal.

Note that, for each $n$, $h$ induces an order $L_n$ on the strings of length $n$: for all $x, y \in \Sigma^n$, $x \preceq_{L_n} y$ if and only if $h(x,y) = y$. (This follows essentially from the observation that $h(x,y) = y$ if and only if $score(x) < score(y)$, where $score$ is defined with respect to $h$. In fact, it is not hard to see that for each $n$ and for each $1 \leq k \leq 2^n$, there is exactly one string of length $n$ having a score of exactly $k$.)

We will below give the procedure *MERGE* that given $S_n$ and $L_{n+1}$ constructs $S_{n+1}$.

Define $S_0 = (\epsilon)$, where $\epsilon$ denotes the empty word. Let $\vdash$ denote the append operation for lists. Now suppose that $S_n = (x_1, x_2, \ldots, x_{2^{n+1}-1})$ and $L_{n+1} = (y_1, y_2, \ldots, y_{2^{n+1}})$ are total orders of the strings in $\Sigma^{*\leq n}$ and $\Sigma^{*n+1}$, respectively.

Procedure **MERGE** [Input $S_n = (x_1, x_2, \ldots, x_{2^{n+1}-1})$, $L_{n+1} = (y_1, y_2, \ldots, y_{2^{n+1}})$.]

$\ell = 1$; $S_{n+1} = $ the empty list;
For $i = 1$ to $2^{n+1} - 1$ do
   If $r = \max\{j \mid \ell \leq j \leq 2^{n+1} \wedge g(x_i, y_j) = x_i\}$ exists do
     $S_{n+1} = S_{n+1} \vdash (y_\ell, y_{\ell+1}, \ldots, y_r, x_i)$;
     $\ell = r + 1$;
   enddo;
   else (i.e., $r$ does not exist) do
     $S_{n+1} = S_{n+1} \vdash (x_i)$;
   enddo;
endfor;
Output $S_{n+1}$;

We have via this procedure completely defined the orders $S_0, S_1, S_2, \ldots$. Clearly, $S_{n+1}$ orders the strings in $\Sigma^{*\leq n+1}$. Furthermore, $S_{n+1}$ respects $S_n$ and also respects $L_{n+1}$.

We will now prove that $f$, defined as $f(x,y) = \max_{S_{\max(|x|,|y|)}}(x,y)$, is a selector for $B$. For each $n$, let $f_n$ be defined as

$$f_n(x,y) = \begin{cases} \max_{S_n}(x,y) & \text{if } |x|, |y| \leq n \\ g(x,y) & \text{otherwise,} \end{cases}$$

for all $x, y \in \Sigma^*$. Note that for all $x, y$, $f(x, y) = f_{\max(|x|,|y|)}(x, y)$. So in order to prove that $f$ is a selector for $B$ it suffices (as the constraints satisfied by selector functions are all defined locally on its two input strings) to show that for all $n$, $f_n$ is a selector of $B$.

It is easy to see that $f_0$ is a selector for $B$. So suppose that $f_i$ is a selector for $B$. Assume that $f_{i+1}$ is not a selector for $B$, in other words—as clearly for all $u$ and $v$, $f_{i+1}(u, v) \in \{u, v\}$— there exist strings $u \in B$ and $v \notin B$ such that $f_{i+1}(u, v) = v$. (As each $f_{i+1}$ is clearly commutative, we know $f_{i+1}(u, v) = f_{i+1}(v, u)$ and so we lose no generality in assuming that $u$ is the string in $B$.) Observe that by the definition of $S_{i+1}$ and the fact that $S_{i+1}$ respects $S_i$ and also respects $L_{i+1}$, we have, for each $x$ and $y$,

$$f_{i+1}(x, y) = \begin{cases} f_i(x, y) & \text{if } |x|, |y| \leq i \\ h(x, y) & \text{if } |x| = |y| = i + 1 \\ \max_{S_{i+1}}(x, y) & \text{if } (|x| = i + 1 \land |y| \leq i) \lor (|x| \leq i \land |y| = i + 1) \\ g(x, y) & \text{otherwise.} \end{cases}$$

Since $f_i$, $h$, and $g$ are selectors for $B$, we must have that either $|u| = i + 1$ and $|v| \leq i$, or $|u| \leq i$ and $|v| = i + 1$. So, by that and our choice of $u$ and $v$, $f_{i+1}(u, v) = \max_{S_{i+1}}(u, v) = v$.

**Case 1 [$|u| = i + 1$ and $|v| \leq i$]:**
By the definition of $S_{i+1}$ (via the MERGE algorithm) this implies that

$$(\exists v' : |v'| \leq i \land v' \preceq_{S_i} v)(\exists u' : |u'| = i + 1 \land u \preceq_{L_{i+1}} u')[g(u', v') = v'].$$

Let $u'$ and $v'$ be such strings. Note that $u \preceq_{L_{i+1}} u'$ implies $h(u, u') = u'$ which in turn, since $u \in B$ and since $h$ is a selector for $B$, implies that $u' \in B$. On the other hand $v' \preceq_{S_i} v$ implies $f_i(v, v') = v$ which in turn, since $v \notin B$ and since $f_i$ is a selector for $B$, implies that $v' \notin B$. Hence $g(u', v') = v'$ contradicts the assumption that $g$ is a selector of $B$.

**Case 2 [$|u| \leq i$ and $|v| = i + 1$]:**
By the definition of $S_{i+1}$ (via the MERGE algorithm) this implies that

$$(\forall u' : |u'| \leq i \land u' \preceq_{S_i} u)(\forall v' : |v'| = i + 1 \land v \preceq_{L_{i+1}} v')[g(u', v') = v'].$$

In particular we have $g(u, v) = v$, a contradiction to the assumption that $g$ is a selector for $B$.

This completes the proof that $f$ is a selector for $B$.

It is not hard to see that $f$ is associative, since for all $x, y, z \in \Sigma^*$, $f(x, f(y, z)) = f(f(x, y), z) = \max_{S_{\max(|x|,|y|,|z|)}}(x, y, z)$. Also, $f$ is clearly commutative.

Regarding the amount of time needed to compute $f$ note that $f(x, y)$ can clearly be computed in time $2^{c \max(|x|,|y|)}$ for some $c$ by simply computing $S_0$ then $S_1$ then ... then $S_{\max(|x|,|y|)}$, and then outputting $\max_{S_{\max(|x|,|y|)}}(x, y)$. This follows in particular from the inductive definition of $S_i$ and the fact that $h$ (and thus each list $L_{i+1}$) can be computed in exponential time. ❑